\documentclass[aip,rsi,reprint]{revtex4-2}
\usepackage{graphicx}
\usepackage{amsmath}
\usepackage[utf8]{inputenx}

\usepackage{dcolumn}
\usepackage{bm}

\usepackage{mathptmx}
\usepackage{etoolbox}

\makeatletter
\def\@email#1#2{%
	\endgroup
	\patchcmd{\titleblock@produce}
	{\frontmatter@RRAPformat}
	{\frontmatter@RRAPformat{\produce@RRAP{*#1\href{mailto:#2}{#2}}}\frontmatter@RRAPformat}
	{}{}
}%
\makeatother

\usepackage{datetime}
\usepackage{color}
\definecolor{rot}{rgb}{1,0,0}
\definecolor{blau}{rgb}{0,0,1}

\usepackage{soul}

\begin{document}

\title{Simultaneous direct measurement of the electrocaloric and dielectric dynamics of ferroelectrics with microsecond temporal resolution}

\author{J. Fischer}
\author{J. D\"{o}ntgen}
\affiliation{Ruhr-Universit\"{a}t Bochum, Fakult\"{a}t f\"{u}r Physik und Astronomie, Arbeitsgruppe Spektroskopie der kondensierten Materie, Universit\"{a}tsstra\ss e 150, 44780 Bochum, Germany}
\author{C. Molin}
\author{S. E. Gebhardt}
\affiliation{Fraunhofer IKTS, Fraunhofer Institute for Ceramic Technologies and Systems, Dresden, Germany}
\author{Y. Hambal}
\author{V. V. Shvartsman}
\author{D. C. Lupascu}
\affiliation{Institute for Materials Science and Center for Nanointegration Duisburg-Essen (CENIDE), University of Duisburg-Essen, 45141 Essen, Germany}
\author{D. H\"{a}gele}
\author{J. Rudolph}
\email{joerg.rudolph@ruhr-uni-bochum.de}
\affiliation{Ruhr-Universit\"{a}t Bochum, Fakult\"{a}t f\"{u}r Physik und Astronomie, Arbeitsgruppe Spektroskopie der kondensierten Materie, Universit\"{a}tsstra\ss e 150, 44780 Bochum, Germany}

\begin{abstract}
A contactless technique for direct time-resolved measurements of the full dynamics of the adiabatic temperature change in electrocaloric materials is introduced. The infrared radiation emitted by the electrocaloric sample is sensitively detected with $\mu$s time resolution and mK temperature resolution. We present time-resolved measurements of the electrocaloric effect up to kHz frequencies of the driving electric field and down to small field strengths. The simultaneous recording of transients for applied electric field and induced polarization gives a comprehensive view on the correlation of electrocaloric and ferroelectric properties. The technique can further be applied to the continuous measurement of fatigue for $>10^6$ electric field cycles.\\
\end{abstract}
\maketitle

%
%
%
		
\section{Introduction}
Cooling technologies are a major contributor to world-wide energy consumption and even further substantial growth is predicted for the next years as the need for air conditioning and refrigeration is expected to strongly increase due to global warming. The development of energy-saving and environmentally friendly sustainable cooling technologies is therefore a key challenge for the near future, as today's cooling technology is still mostly based on gas expansion as already patented by Carl von Linde in 1877. Promising candidates for improved cooling technologies are caloric effects in solids, where the application or removal of an electric, magnetic or mechanical field leads to a reversible adiabatic temperature change $\Delta T_{\mathrm{ad}}$ and an isothermal entropy change $\Delta S$, respectively. \cite{Kitanovski_IntJRefrig_57_288,Faehler_AdvEngMater_14_10} The corresponding electrocaloric, magnetocaloric, and elasto- or barocaloric effects have attracted strongly growing interest in the last years.\cite{Moya_Science_370_797,Moya_NatureMater_13_439} Pioneering work on caloric effects had been mostly concentrated on the magnetocaloric effect, while the electrocaloric effect (ECE) had been considered to be too small for applications for a long time after its early demonstration in 1930.\cite{Kobeko_ZPhys_66_192} The discovery of giant electrocaloric effects in thin films\cite{Mischenko_Science_311_1270,Neese_Science_321_821} has, however, triggered intense research activities on both fundamentals and applications of the ECE.\cite{Torello_AdvElectronMater_8_2101031} The electrocaloric effect has the important benefit of being technologically very attractive as high electric fields are considerably easier to generate than high magnetic fields.

A mandatory prerequisite for fundamental investigations of the ECE as well as for the characterization of novel electrocaloric materials is the reliable and precise determination of the adiabatic temperature change $\Delta T_{\mathrm{ad}}$ which is generally possible via indirect methods or via direct measurements of the field-induced temperature change. The large majority of published experimental data on the electrocaloric effect has been retrieved by indirect methods,\cite{Liu_ApplPhysRev_3_0311022,Chen_APL_118_122904,Cheng_PhysStatusSolidiA_216_1900684} which mostly rely on experimentally easily accessible measurements of the field and temperature dependent polarization $P(T,E)$. Integrating the relevant Maxwell relation yields values for $\Delta S$, from which $\Delta T_{\mathrm{ad}}$ is estimated via $\Delta T_{\mathrm{ad}}\approx -T\Delta S/c$ with the heat capacity $c$.\cite{Cheng_PhysStatusSolidiA_216_1900684,Kutnjak_inElectrocaloricMaterials,Liu_ApplPhysRev_3_0311022,Moya_NatureMater_13_439} Indirect measurements are, however, prone to inaccuracies and can result in substantial deviations from directly determined values, especially in the case of relaxor ferroelectrics.\cite{Sanlialp_APL_106_062901,Sanlialp_APL_111_173903,LeGoupil_APL_107_172903,Chen_APL_118_122904,Lu_APL_97_202901,Birks_JAP_121_224102} The direct determination of the adiabatic temperature change $\Delta T_{\mathrm{ad}}$ might seem to be conceptually simple at first glance, but its practical realization is hindered by several requirements, which can be only approximately fulfilled. Most importantly, the temperature change $\Delta T_{\mathrm{ad}}$ has to be measured under adiabatic conditions. However, the sample temperature has to be set and controlled in an experiment, which is usually achieved by thermally coupling the sample to a temperature-controlled sample holder, thus contradicting adiabaticity. In addition, the temperature change induced by the electrocaloric effect is often measured by a temperature sensor like a thermocouple or a resistive sensor, which requires thermal equilibration with the sample through heat exchange, thus again contradicting adiabaticity. As a consequence, the mass of the temperature sensor should be as small as possible, which, in turn, makes the fabrication of the sensors technologically very challenging and causes additional problems like reliably establishing a good thermal contact between sample and temperature sensor.\cite{Liu_ApplPhysRev_3_0311022}

These problems get even more severe for the study of thin-film samples, where the mass and response time of temperature sensors would have to be extremely small. In addition, thin-film samples are usually prepared on an underlying substrate. To prevent rapid heat exchange with the substrate, the temperature measurement has to be significantly faster than the time required for thermal equilibration of thin-film and substrate. Such fast measurements are almost impossible with temperature sensors like thermocouples or thermistors, with technologically very complex exceptions like sophisticated sample designs with specifically deposited thin-film resistive temperature sensors\cite{Pandya_PRAppl_7_034025} or thermocouples.\cite{Matsushita_ApplPhysExpr_13_041007,Matsushita_JpnJApplPhys_55_10TB04}

One approach to circumvent the problems sketched above is to measure the sample temperature by contactless methods, e.g. via the infrared (IR) radiation emitted by the sample, either using IR detectors\cite{Lu_APL_97_162904,Sotnikova_RSI_91_015119} or IR cameras.\cite{KarNarayan_APL_102_032903,Sebald_APL_101_022907,Guo_APL_105_031906,Nouchokgwe_NatureCommunications_12_3298,Nair_Nature_575_468} While such optical techniques allow in principle for large detection bandwidths, the bandwidth has been limited to approximately $100$~Hz so far and the dynamics $\Delta T_{\mathrm{ad}}(t)$ of the electrocaloric effect has not been studied. 

Systematic studies of the dynamics of the electrocaloric effect are, however, highly desirable. Such investigations would on the one hand allow for a deeper understanding of the fundamentals of the ECE: the role of, e.g., the relaxational behavior of the polarization\cite{Rowley_JPhysCondensMatter_27_395901} or of the complex dynamics caused by switching and growth of domains\cite{Cheng_PhysStatusSolidiA_216_1900684,Liu_ApplPhysRev_3_0311022} has so far not been systematically studied. The correct interpretation of experimental results, on the other hand, also strongly relies on a sound knowledge of the dynamics of the observed temperature changes: the temperature change $\Delta T_{\mathrm{ex}}$ which is observed in an experiment, will in general not be identical to the ideal adiabatic temperature change $\Delta T_{\mathrm{ad}}$, but will also include contributions from, e.g., Joule heating or heat exchange with the surrounding.\cite{Bradesko_APLMater_7_071111,Quintero_APL_99_232908}

Here, we present a technique for direct, contactless measurements of the full dynamics of the electrocaloric effect with mK temperature resolution and sub-ms temporal resolution. The direct measurement of the experimentally observed temperature change $\Delta T_\mathrm{ex}(t)$ with its dynamics is based on a large-bandwidth detection of the IR radiation emitted by the sample. \cite{Doentgen_APL_106_032408,Doentgen_RevSciInstrum_89_033909,Doentgen_EnergyTechnol_6_1470} Simultaneously measured transients of the temperature change $\Delta T_\mathrm{ex}(t)$, the driving electric field $E(t)$ and the induced polarization $P(t)$ allow for a comprehensive characterization of the dynamics of the electrocaloric effect and the correlation with dielectric properties. The technique can also be used to quasi-continuously monitor fatigue of samples over many cycles of the driving electric field. The technique is suitable for ferroelectric samples with thicknesses ranging from millimeters down to several micrometers without the need for complex sample preparation.

In the following sections, we will first introduce the basic concept for the direct, time-resolved measurement of the temperature change $\Delta T_{\mathrm{ex}}(t)$ of the sample, before the individual components of the setup and the temperature calibration routine will be discussed in detail. Finally, the technique is demonstrated for different ferroelectric material systems ranging from a bulk sample to thin polymer films.

\section{Basic concept}
\begin{figure*}
	\centering
	\includegraphics[width=1.7\columnwidth]{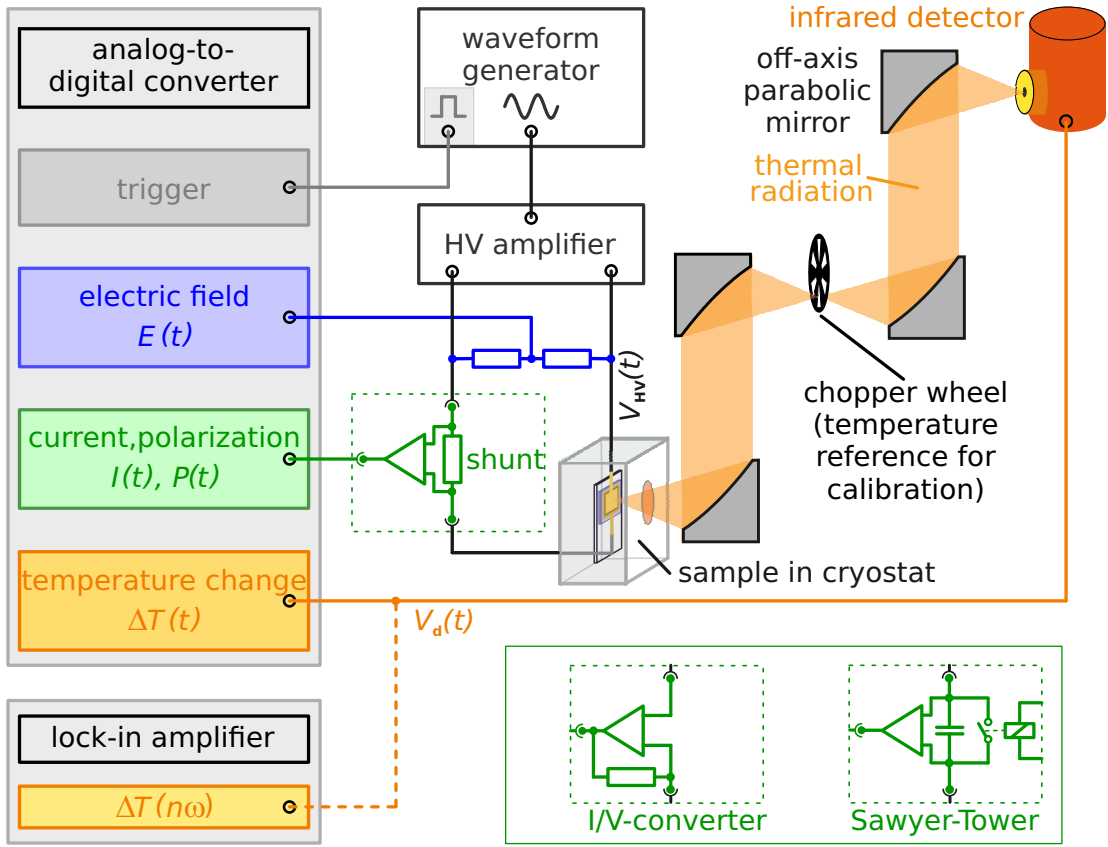}
	\caption{(Color online) Schematic setup for the direct measurement of the temperature change $\Delta T_{\mathrm{ex}}(t)$ of an electrocaloric sample via the infrared radiation emitted by the sample. The electric field $E(t)$ and resulting current $I(t)$ or polarization $P(t)$ are measured simultaneously by an analog-to-digital converter. The inset shows circuits for measurement of current via a transimpedance amplifier ($I/V$-converter) and a Sawyer-Tower circuit.}
	\label{Fig:Setup}
\end{figure*}

Our setup for non-contact direct, time-resolved measurements of the temperature change $\Delta T_\mathrm{ex}(t)$ is schematically shown in Figure~\ref{Fig:Setup}. The electrocaloric sample is mounted in vacuum in a cold-finger cryostat on a specifically designed sample holder. A time-dependent high voltage $V_{\mathrm{HV}}(t)$ is generated by a versatile waveform generator with subsequent high-voltage amplifier, and is applied to the sample via a small electrical network which allows for simultaneous dielectric measurements. The corresponding electric field $E(t)=V_{\mathrm{HV}}(t)/d$ for a sample with thickness $d$ leads to temperature changes $\Delta T_\mathrm{ex}(t)$ of the sample due to the electrocaloric effect and corresponding changes of the emitted IR radiation. The IR radiation is focused onto a liquid-nitrogen cooled HgCdTe-IR-detector via two pairs of off-axis parabolic mirrors. The detector's output signal $V_{\mathrm{d}}(t)$ is recorded by a multi-channel analog-to-digital converter, which simultaneously also captures the applied electric field $E(t)$ as well as the resulting current $I(t)$ or polarization $P(t)$. Absolute values for the sample temperature change are obtained via a separate calibration measurement, where an optical chopper wheel serves as a constant temperature reference. 

\section{Experimental setup}
In the following, we will discuss the key components of the setup in detail.
\subsection{Generation of the driving high voltage}
A time-dependent high voltage $V_{\mathrm{HV}}(t)$ is generated by the combination of a versatile waveform generator and a high-voltage amplifier. For fundamental measurements we typically use a bipolar, sinusoidally oscillating voltage $V_{\mathrm{HV}}(t)=V_{\mathrm{p}}\sin (\omega t)$ with amplitude $V_p$ and frequency $\omega = 2\pi f$, but other waveforms like a unipolar sine wave or triangular waveforms for constant field-change rates can be easily applied in a straightforward way (see Section~\ref{Sec:BZTsample}). Depending on the required driving voltage and current either a high-voltage amplifier with maximum voltage of $150$~V capable of delivering currents up to $300$~mA (Falco Systems, model WMA-300) or an amplifier with maximum voltage of $7.5$~kV and currents up to $50$~mA (TReK Inc., model PD05034) is used, respectively.

A trigger signal can be generated synchronized to the desired waveform to allow for triggered data acquisition in single-shot measurements, where, e.g., bursts with predefined number of cycles of the electric field are used.

\subsection{\label{Sec:SampleHolder}Optical cryostat and sample holder}
\begin{figure*}
	\centering
	\includegraphics[width=2.0\columnwidth]{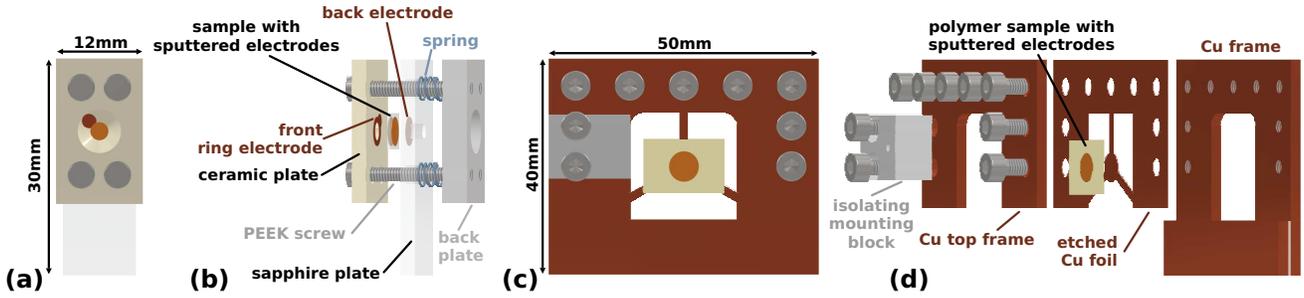}
	\caption{(Color online) (a) Front view and (b) side view of the spring-loaded sample holder. The use of soft springs with small spring constant allows to carefully control the clamping force when the screws are tightened to fix the sample. The sample holder is mounted via the sapphire plate to the copper cold-finger of the cryostat. (c) Front view and (d) tilted side view of the sample holder for thin film samples: the thin film sample is mounted on a thin copper foil which is fixed to a copper frame. Suspending the sample via three narrow legs of the copper foil minimizes heat exchange with the sample holder.}
	\label{Fig:SampleHolder}
\end{figure*}

The sample is mounted in vacuum in a custom-built cold-finger cryostat with optical access through an anti-reflection coated ZnSe window which possesses high transmission for IR radiation and visible light, thus allowing for a convenient visual alignment of the setup. The sample chamber is evacuated by a turbo-molecular vacuum pump to a base pressure of typically $<10^{-5}$~mbar to improve the adiabaticity of the measurements and to prevent electrical breakdown of the residual gas in the sample chamber. The temperature of the copper cold-finger is controlled purely electrically via two stacked high-temperature Peltier elements which are mounted on a water-cooled cold plate, allowing for a temperature range approximately from $250$~K to $420$~K. Different sample holders can be mounted to the cold-finger. In most measurements, a sample with sputter-deposited electrodes is mounted onto a spring-loaded sample holder [cf. Fig.~\ref{Fig:SampleHolder}(a) and (b)], where the front electrode of the sample is contacted via a copper ring-electrode held by a glass ceramic plate. The central aperture of the ring electrode is substantially larger than the size of the IR-detector, thus allowing for full optical access to the sample. The glass ceramic plate with the front ring-electrode is gently pressed against the sample by four springs, which allow in combination with four tightening screws to carefully control the clamping force. The spring constant is chosen to be small to avoid spurious effects due to unwanted strain in the sample. The tightening screws are made from PEEK which resists high voltages and can be used in a wide temperature range. The back-electrode of the sample is contacted via a circular copper electrode mounted on a $30$~mm by $10$~mm large and $2$~mm-thick sapphire plate which is attached to the copper cold-finger, thus allowing for a good control of the sample temperature while electrically isolating the high-voltage applied to the sample from the cryostat. All materials of the sample holder are chosen to be compatible with high temperatures and high voltages. The base temperature of the sample is monitored via a small PT100 resistive temperature sensor attached to the surface of the sample or to the sapphire plate. While this sample holder has the advantage of good control over the sample temperature and an easy and fast exchange of samples, measurements have to be performed fast enough to achieve adiabaticity (see discussion of $\Delta T$ measurements for different driving frequencies in Section~\ref{Sec:BZTsample}).

Thin samples like polymer films (cf. Section~\ref{Sec:Polymerfilms}) are usually mounted to a thin, free standing copper foil by an electrically conductive epoxy to minimize heat exchange with the sample holder. Typically, a $60$~$\mu$m-thick Cu foil is structured by photolithography and wet chemical etching so that the sputter-deposited electrode of the sample is fully contacted and the sample is attached by a spider leg configuration to an outer copper frame [cf. Fig.~\ref{Fig:SampleHolder}(c) and (d)]. The width of the spider legs can be varied to adjust thermal coupling to the outer copper frame, which is mounted to the cold-finger of the cryostat for temperature control. The copper frame and the copper foil act as the electrical back contact to the sample and are electrically isolated from the cryostat by sapphire plates. The front electrode of the sample is contacted via a thin copper wire glued to the sputter-deposited electrode with an electrically conductive epoxy and mechanically anchored at an isolating mounting block on the copper frame. 

\subsection{Optical beam path}
The IR radiation emitted by the sample is focused onto the IR detector by two pairs of 90$^\circ$ off-axis parabolic mirrors in a telescopic arrangement. The off-axis parabolic mirrors with a diameter of $101.6$~mm and an effective focal length of $152.4$~mm have an enhanced aluminum coating for high IR reflectivity. The liquid-nitrogen cooled photoconductive HgCdTe-detector (InfraRed associates Inc., model MCT-13-0.50) has a size of $500\times 500~\mu\mathrm{m}^2$ with a peak detectivity at a wavelength of approximately $12~\mu\mathrm{m}$ and a cut-off wavelength of approximately $13~\mu\mathrm{m}$. The detector output signal is amplified by a specifically matched amplifier with a high-pass filter with a cut-off frequency of $1.5$~Hz in the output. The bandwidth of the IR detector and preamplifier extends to $150$~kHz. An optical chopper wheel is placed at the intermediate focus of the beam path to serve as a constant-temperature reference thermal emitter in separate calibration measurements for the observed temperature change (see Section~\ref{Sec:DeltaTcalibration}).

\subsection{{\label{SecElectrSetup}}Electrical setup for simultaneous dielectric measurements}
The dielectric properties of the caloric sample are measured simultaneously with the temperature change $\Delta T_{\mathrm{ex}}(t)$ of the sample. For this purpose, the driving high-voltage $V_{\mathrm{HV}}(t)$ is applied to the sample via a small network. The applied voltage is measured in one branch of the network with an Ohmic voltage divider (see blue resistors in Fig.~\ref{Fig:Setup}) built from thick-film resistors rated for high voltage, where the voltage drop across the sensing resistor is buffered via a high input impedance unity-gain amplifier. The polarization $P(t)$ of the sample is preferentially determined via the current $I(t)$ flowing through the sample by numerical integration $P(t) = [(1/A)\int_0^t I(\tau) \mathrm{d}\tau] - \varepsilon_0E(t)$ with $A$ as the contact area. The current $I(t)$ is measured in the other branch of the network either via the voltage drop across a low-resistance shunt resistor buffered by a low offset-voltage amplifier, or via a custom-built transimpedance amplifier (see green circuit part in Fig.~\ref{Fig:Setup}). Alternatively to current measurements, the polarization $P(t)$ can also be detected by a Sawyer-Tower circuit.\cite{Sawyer_PR_35_269} In this case, a low-loss mica capacitor is chosen for the reference capacitor, with a capacitance at least two orders of magnitude larger than the capacitance of the sample. The reference capacitor is automatically discharged by the control software via a relay just before the measurement, and the voltage across the reference capacitor is buffered via a high-input impedance operational amplifier. 

The circuits for current measurements via a shunt and via a transimpedance amplifier, or for polarization measurements by the Sawyer-Tower method can be easily interchanged due to a modular design. The results of the different methods were checked for consistency. We have further checked that no instrumental phase shifts occur between the IR detector signal and the signals for $E(t)$, $I(t)$ and $P(t)$.

The dielectric permittivity $\varepsilon (T)$ of samples is conveniently measured within the same setup by omitting the high-voltage amplifier and directly applying a small sinusoidal voltage to the dielectric measurement network [cf. Fig.~\ref{Fig:DielectricCharacterization}(a)]. The corresponding output voltages for electric field and induced current (or polarization if using the Sawyer-Tower circuit) are in this case detected by lock-in amplifiers. The results were checked for consistency with measurements of the relative permittivity performed with a precision $LCR$ meter (HP4284A) and showed excellent agreement.

\subsection{\label{Sec:dataacquisition}Data acquisition}
The IR detector signal $V_\mathrm{d}(t)$ as well as the output signals of the dielectric characterization circuit for electric field $E(t)$ and current $I(t)$ or polarization $P(t)$ are simultaneously digitized by a multi-channel analog-to-digital converter (Pico Technology Ltd., model PicoScope 4444). The analog-to-digital converter (ADC) is controlled via a Python program that allows to average multiple measurements with a predefined number of samples and sample rate, or to record triggered single-shot experiments. The optional use of a lock-in amplifier (see Figure~\ref{Fig:Setup}) allows either for a quasi-continuous detection of the dominant harmonic contribution to the IR signal, e.g. for measurements of fatigue (see Section~\ref{Sec:Fatigue}), or for the detection of very weak signals. In this case, the measurement of the signal at several harmonics $n\omega$ for a driving field with frequency $\omega$ still allows to retrieve the dynamics of the electrocaloric effect via Fourier reconstruction.\cite{Doentgen_RevSciInstrum_89_033909}

\subsection{\label{Sec:DeltaTcalibration}Calibration of $\Delta T_{\mathrm{ex}}$}
\begin{figure}
	\centering
	\includegraphics[width=1.0\columnwidth]{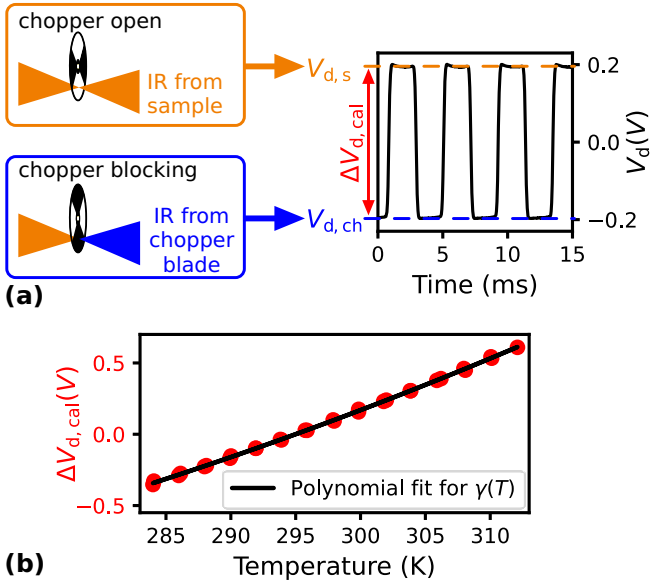}
	\caption{(Color online) (a) Procedure for the calibration of the sample temperature change. A spinning chopper blade in the intermediate focal plane of the beam path serves as a constant temperature reference while the sample temperature is swept. The detector detects either the IR radiation from the sample when the chopper blade is open, or it detects the IR radiation emitted by the chopper blade when the chopper blade blocks the beam path. (b) Typical example of the difference $\Delta V_{\mathrm{d,cal}}$ between the signal levels $V_{\mathrm{d,s}}$ and $V_{\mathrm{d,ch}}$ for the two cases as a function of the sample temperature $T$. A second-order polynomial fit $\gamma(T)$ to $\Delta V_{\mathrm{d,cal}}(T)$ allows to retrieve a calibration function for the calibration of $\Delta T_{\mathrm{ex}}$.}
	\label{Fig:Calibration}
\end{figure}
The sample temperature change $\Delta T_{\mathrm{ex}}(t)$ is obtained from the IR detector signal $V_{\mathrm{d}}(t)$ via a separate calibration measurement due to which the emissivity of the sample has not to be known explicitely. In principle, the calibration corresponds to recording the detector output signal as a function of the sample temperature. However, we use a modulation scheme for the calibration, where a rotating chopper wheel in the intermediate focal plane of the IR beam path serves as a constant temperature reference.\cite{Doentgen_RevSciInstrum_89_033909} In this way, the calibration signal is shifted to frequencies above the cut-off frequency of the high-pass filter of the IR detector amplifier, and the calibration becomes less susceptible to slow drifts, e.g., of background thermal radiation. The IR detector output in this case approximately is a square wave oscillating with the chopping frequency $\omega_{\mathrm{ch}}$ between the signal levels $V_{\mathrm{d,s}}(T)$ for the IR intensity emitted by the sample with temperature $T$ and $V_{\mathrm{d,ch}}(T_{\mathrm{ch}})$ for the IR intensity emitted by the chopper blade with constant temperature $T_{\mathrm{ch}}$ [cf. Figure~\ref{Fig:Calibration}(a)]. The signal difference $\Delta V_{\mathrm{d,cal}}(T) = V_{\mathrm{d,s}}(T) - V_{\mathrm{d,ch}}(T_{\mathrm{ch}})$ between these two levels is recorded as a function of the simultaneously measured sample temperature $T$, giving $\Delta V_{\mathrm{d,cal}}(T) = \gamma(T)$. We typically use a second-order polynomial fit to obtain $\gamma(T)$ from $\Delta V_{\mathrm{d,cal}}(T)$ [see Figure~\ref{Fig:Calibration}(b)]. Then, a change $\Delta T$ of the sample temperature results in a corresponding change $\Delta V_{\mathrm{d}}$ of the detector signal
\begin{equation}
\Delta V_{\mathrm{d}} = \left(\frac{\mathrm{d}\gamma(T)}{\mathrm{d}T}\right) \Delta T
\end{equation}
with $\mathrm{d}\gamma(T)/\mathrm{d}T$ as the derivative of the second-order polynomial. This calibration measurement is performed for each individual sample to account for the individual emissivities of different samples. While this procedure requires the extra effort of additional calibration measurements, it has the benefit that the change of the detector signal is a direct function of the sample temperature change and the emissivity of the sample is included only implicitly in the detector signal, but has not to be determined separately nor an assumption about the emissivity has to be made.

The calibrated sample temperature change $\Delta T_{\mathrm{ex}}(t)$ is finally obtained as
\begin{equation}
	\Delta T_{\mathrm{ex}}(t) = \Delta T_{\mathrm{ex,AC}}(t) + \Delta T_{\mathrm{ex,DC}} = \left(\frac{\mathrm{d}\gamma}{\mathrm{d}T}\right)^{-1} \Delta V_{\mathrm{d}}(t) + \Delta T_{\mathrm{ex,DC}} \ .
\end{equation}
The extra term $\Delta T_{\mathrm{ex,DC}}$ of the temperature change has to be added to compensate for the effect of the high-pass filter in the IR detector output, where the correct $0 \leq \Delta T_{\mathrm{ex}}(t) \leq \Delta T_{\mathrm{ex}}^{\mathrm{max}}$ is retrieved for $\Delta T_{\mathrm{ex,DC}} = |\mathrm{min}(\Delta T_{\mathrm{ex,AC}})|$ in the case of the normal electrocaloric effect. Alternatively, the DC term can also be retrieved from burst measurements, where the electrocaloric effect is measured in a triggered acquisition for a finite number $N$ of electric field cycles. We have found it necessary to perform the calibration on a sample spot which was coated with a matt black paint suitable for high temperatures (Electrolube, PNM400). The black paint increases the emissivity of the sample as compared to the bare metallic electrodes. The increased emissivity makes the calibration less prone to artifacts, e.g., from stray IR radiation. We have coated the chopper blade with the same matt black paint to achieve similar emissivities of sample and chopper, which results in a zero-crossing of $\Delta V_{\mathrm{d,cal}}$ at room-temperature [cf. Figure~\ref{Fig:Calibration}(b)]. The higher emissivity on a sample spot coated with black paint also leads to a better signal-to-noise ratio for measurements of the sample temperature change $\Delta T_{\mathrm{ex}}$. However, the additional layer of black paint can distort the dynamics $\Delta T_{\mathrm{ex}}(t)$ of the sample temperature change for high frequencies (see Section~\ref{Sec:PMN8PT})). In such a case, we perform measurements of the dynamics of $\Delta T_{\mathrm{ex}}$ on the bare electrode without a layer of black paint, and match the calibration to the calibration on a matt black painted sample spot at low frequency. 

\section{Dynamics of the electrocaloric effect in ferroelectrics}
In this section, we demonstrate our technique by presenting time-resolved measurements of the dynamics of the temperature change $\Delta T_{\mathrm{ex}}(t)$ together with simultaneous measurements of the dielectric properties in several ferroelectric samples. We will discuss measurements of the electrocaloric effect in a bulk Ba(Zr$_{0.12}$Ti$_{0.88}$)O$_3$ ceramic as a first example, before we show measurements on relaxor PMN-PT films and PVDF polymer films as examples for film samples. 
\subsection{\label{Sec:BZTsample}Ba(Zr$_{0.12}$Ti$_{0.88}$)O$_3$ bulk sample}
\begin{figure}
	\centering
	\includegraphics[width=1.0\columnwidth]{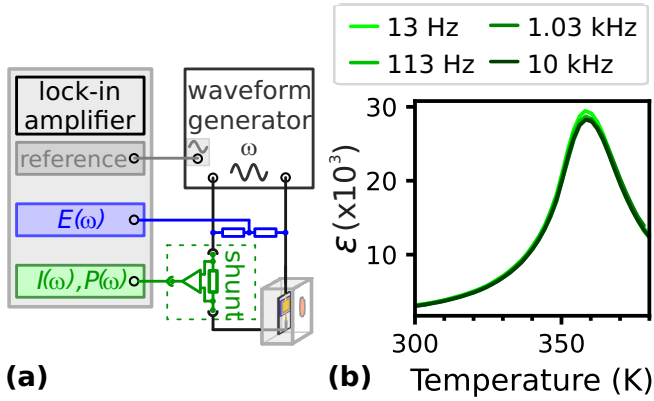}
	\caption{(Color online) (a) Setup for dielectric characterization and (b) example measurement of the dielectric permittiviy $\varepsilon (T)$ for different driving frequencies for a bulk BZT-12 sample.}
	\label{Fig:DielectricCharacterization}
\end{figure}
The Ba(Zr$_x$Ti$_{1-x}$)O$_3$ (BZT) material system has attracted interest as a lead-free electrocaloric material.\cite{Qian_AdvFuncMater_24_1300,Ma_JAP_121_024103,Ye_APL_105_152908} The Zr content allows to tune the transition temperature and to change BZT to a relaxor ferroelectric for high Zr contents $x> 0.2$. Here, we investigate a bulk ceramic Ba(Zr$_{0.12}$Ti$_{0.88}$)O$_3$ (BZT-12) sample which was synthesized using a mixed oxide route.\cite{Ma_JAP_121_024103} The disk-shaped sample with a thickness of $1$~mm and a diameter of approximately $8$~mm has sputter-deposited Au electrodes on both sides and was mounted on the spring-loaded sample holder discussed in Section~\ref{Sec:SampleHolder}. The temperature dependence $\epsilon(T)$ of the real part of the relative permittivity was measured as described in Section~\ref{SecElectrSetup} and shows a broad peak with maximum $\epsilon \approx 30\times 10^3$ at a temperature of approximately $358$~K [cf. Figure~\ref{Fig:DielectricCharacterization}(b)], in good agreement with published data for BZT-12.\cite{Sanlialp_IEEETransUltrasonFerroelecFreqControl_63_1690} 

\begin{figure}
	\centering
	\includegraphics[width=1.0\columnwidth]{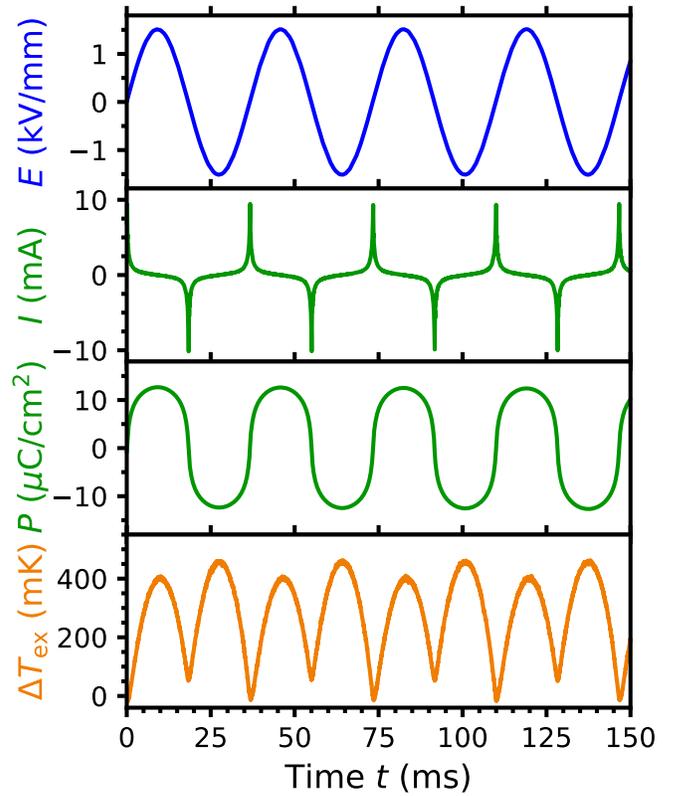}
	\caption{(Color online) Transients of electric field $E(t)$, current $I(t)$, polarization $P(t)$, and temperature change $\Delta T_\mathrm{ex}(t)$ (from top to bottom) for the bulk BZT-12 sample at a temperature of $358$~K for a driving electric field with an amplitude of $1.5$~kV/mm and a frequency of $27$~Hz.}
	\label{Fig:Transients}
\end{figure}
\begin{figure}
	\centering
	\includegraphics[width=1.0\columnwidth]{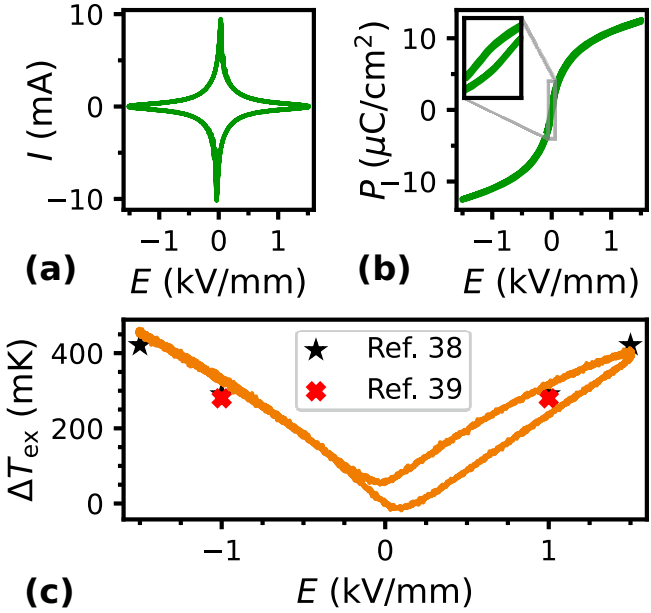}
	\caption{(Color online) Electric field dependence of (a) current $I(E)$, (b) polarization $P(E)$ and (c) temperature change $\Delta T_{\mathrm{ex}}(E)$ for the bulk BZT-12 sample obtained from the corresponding transients at a temperature of $358$~K for a driving electric field with an amplitude of $1.5$~kV/mm and a frequency of $27$~Hz. For comparison with literature, the symbols in (c) show the adiabatic temperature change $\Delta T_{\mathrm{ad}}$ for electric fields of $1$~kV/mm and $1.5$~kV/mm taken from Ref.~\onlinecite{Sanlialp_IEEETransUltrasonFerroelecFreqControl_63_1690} and \onlinecite{Sanlialp_RSI_89_034903}.}
	\label{Fig:EFieldDependencies}
\end{figure}

Figure~\ref{Fig:Transients} exemplarily shows transients of the applied electric field $E(t)$, the resulting current $I(t)$ through the sample and the detected temperature change $\Delta T_{\mathrm{ex}}(t)$ of the sample at a base temperature of $358$~K and a driving frequency of $27.3$~Hz. The temporal evolution of the polarization $P(t)$ additionally shown in Figure~\ref{Fig:Transients} is obtained from the current transient $I(t)$ by numerical integration. Most importantly, the transient $\Delta T_{\mathrm{ex}}(t)$ directly shows the full dynamics of the sample temperature change induced by the driving electric field $E(t)$. The mK temperature resolution and sub-ms time resolution together with the very good signal-to-noise ratio of $\Delta T_{\mathrm{ex}}(t)$ clearly illustrate the benefits of our technique. Several features are immediately visible from the transients: First, the temperature change of the sample approximately follows the absolute value of the driving electric field, leading to temperature oscillations at twice the frequency $2f$ of the applied electric field with frequency $f$ as is expected for a bipolar driving field. Second, the ECE shows slightly different amplitudes for positive and negative electric fields. This asymmetry is also found in the different amplitudes of the switching peaks in the transients of current and might be caused by fields due to defects. Third, the $\Delta T_\mathrm{ex}(t)$ transients for each field direction posses approximate time-inversion symmetry, thus indicating only very weak irreversible processes\cite{Doentgen_EnergyTechnol_6_1470} in accordance with the very narrow hysteresis loop of BZT (see also below). Fourth, the maximum temperature change $\Delta T_{\mathrm{ex}}^{\mathrm{max}} = \max\left[\Delta T_{\mathrm{ex}}(t)\right]\approx 450$~mK for a field of $1.5$~kV/mm is in very good agreement with reported values in literature [see also Figure~\ref{Fig:EFieldDependencies}(c)].\cite{Sanlialp_RSI_89_034903,Sanlialp_IEEETransUltrasonFerroelecFreqControl_63_1690,Ma_JAP_121_024103}

Combining the transients for current $I(t)$ and polarization $P(t)$ with the electric field transient $E(t)$ further directly gives the electric field dependent current $I(E)$ and the electric field dependent polarization $P(E)$ shown in Fig.~\ref{Fig:EFieldDependencies}. The electric field dependent current $I(E)$ diagram clearly shows typical switching peaks, and the electric field dependent polarization $P(E)$ gives the typical loop with small hysteresis, which is in good agreement with published data for BZT-12.\cite{Sanlialp_IEEETransUltrasonFerroelecFreqControl_63_1690} Combining the transients for temperature change $\Delta T_{\mathrm{ex}}(t)$ and electric field $E(t)$ allows to obtain the full field dependence $\Delta T_{\mathrm{ex}}(E)$ of the ECE from a single measurement [see Fig.~\ref{Fig:EFieldDependencies}(c)]. The mK temperature resolution of our technique allows to track the field dependence down to very small fields. We attribute the asymmetry of $\Delta T_{\mathrm{ex}}(E)$ with a loop-like structure for positive field direction visible in Figure~\ref{Fig:EFieldDependencies}(c) to an asymmetry of the field-dependent current, resulting in increased Joule heating for positive field direction.

\begin{figure}
	\centering
	\includegraphics[width=1.0\columnwidth]{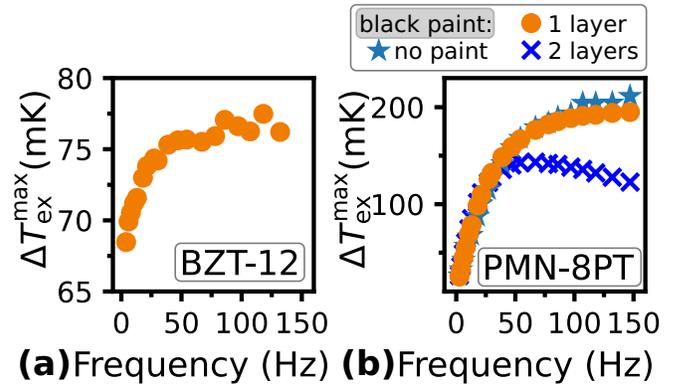}
	\caption{(Color online) (a) Frequency dependence of the maximum temperature change $\Delta T_{\mathrm{ex}}^{\mathrm{max}}$ for a $1$~mm thick BZT-12 bulk ceramic sample for an electric field amplitude of $0.25$~kV/mm. (b) Frequency dependence of the experimentally observed temperature change $\Delta T_{\mathrm{ex}}^{\mathrm{max}}$ for a $38$~$\mu$m PMN-8PT thick film on an alumina substrate for an electric field amplitude of $3.9$~kV/mm. The measurements are taken on sample spots which are uncoated (stars), coated with a single, $5$~$\mu$m thick layer (circles), and two layers with total thickness of $10$~$\mu$m (crosses) of black paint.}
	\label{Fig:DeltaTvsFrequency}
\end{figure}
For systematic studies of the dynamics of the electrocaloric effect, the frequency $f$ of the driving field can be varied to investigate the influence on $\Delta T_{\mathrm{ex}}$ of, e.g., extrinsic contributions like heat exchange with the surrounding or of the intrinsic dynamics of the domain system. Figure~\ref{Fig:DeltaTvsFrequency}(a) shows the experimentally observed maximum temperature change $\Delta T_{\mathrm{ex}}^{\mathrm{max}}$ for frequencies from $2.7$~Hz up to $147$~Hz at $358$~K for a field amplitude of $0.25$~kV/mm.\cite{Note1} The lower electric field as compared to the transients shown in Figure~\ref{Fig:Transients} has been chosen here and for all following measurements to minimize effects of fatigue. The temperature change shows a very steep increase at low frequencies followed by a weak, approximately linear increase for higher frequencies. The steep increase at low frequencies can be well understood by increasing adiabaticity: with increasing driving frequency the temperature change in the sample gets faster than the thermalization with the sample holder. The slower increase for higher frequencies might still be caused by heat exchange with the surrounding or by a weak frequency dependence of the ECE, but more detailed studies are needed to clarify this point.

\begin{figure}
	\centering
\includegraphics[width=1.0\columnwidth]{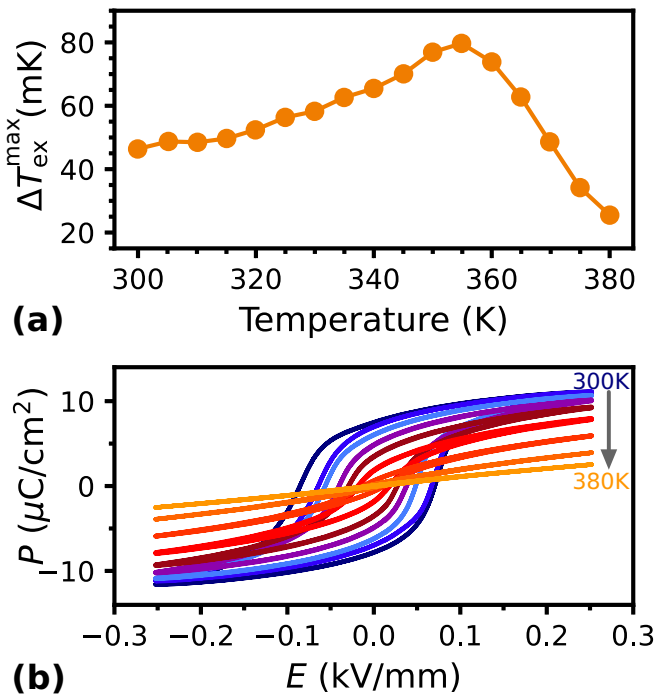}
	\caption{(Color online) (a) Electrocaloric effect in bulk BZT-12 as a function of sample temperature $T$ for a driving electric field with an amplitude of $0.25$~kV/mm and a frequency of $27$~Hz. (b) $P(E)$ loops measured simultaneously with the temperature change $\Delta T_{\mathrm{ex}}(T)$ shown in (a). The $P(E)$ loops are shown in steps of $10$~K for clarity.}
	\label{Fig:BZT12DeltaTvsT}
\end{figure}
The temperature dependence of the ECE can be obtained by extracting the maximum temperature change $\Delta T_{\mathrm{ex}}^{\mathrm{max}}$ for different sample base temperatures $T$. Figure~\ref{Fig:BZT12DeltaTvsT}(a) shows the temperature dependence of the ECE for the bulk BZT-12 sample as an example, which is in good agreement with literature.\cite{Ma_JAP_121_024103,Sanlialp_RSI_89_034903} Figure~\ref{Fig:BZT12DeltaTvsT}(b) shows corresponding $P(E)$-loops for different sample temperatures, which are obtained simultaneously with the temperature-dependent measurements of $\Delta T_{\mathrm{ex}}$. Such combined measurements of caloric and polarization properties allow for a comprehensive characterization of caloric materials under identical conditions. In contrast, the electrocaloric and polarization response have usually been investigated separately for dinstinctively different frequencies and waveforms of the driving electric field so far in literature.

\begin{figure*}
	\centering
	\includegraphics[width=1.7\columnwidth]{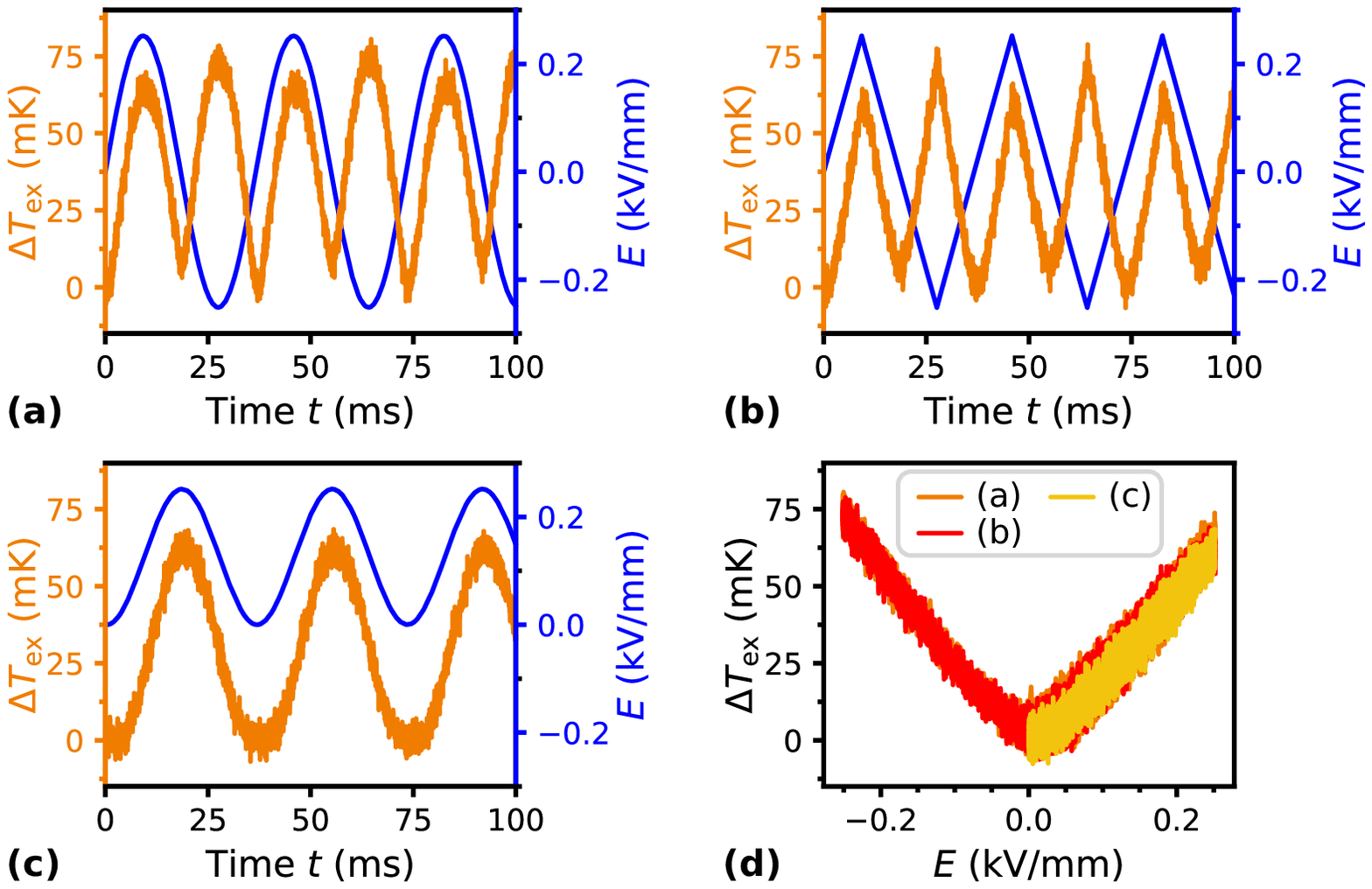}
	\caption{(Color online) Transients of the temperature change $\Delta T_{\mathrm{ex}}(t)$ for bulk BZT-12 for (a) a bipolar sinusoidal, (b) a bipolar triangular and (c) a unipolar sinusoidal driving electric field with an amplitude of $0.25$~kV/mm and a frequency of $f=27.3$~Hz at a temperature of $359$~K. (d) Comparison of the field-dependence $\Delta T_{\mathrm{ex}}(E)$ for the three electric field waveforms shown in (a)-(c).}
	\label{Fig:DifferentDrivingWaveforms}
\end{figure*}

The use of a versatile waveform generator for the generation of the time-dependent driving electric field $E(t)$ allows to conveniently study the influence of different waveforms for $E(t)$ on the ECE. Figure~\ref{Fig:DifferentDrivingWaveforms} compares transients $\Delta T_{\mathrm{ex}}(t)$ for a bipolar sinusoidal and triangular driving field as well as for a unipolar sinusoidal field. It can be clearly seen that the temperature change shows the expected behavior: for the bipolar driving fields with frequency $f$, the temperature change $\Delta T_{\mathrm{ex}}(t)$ follows the absolute value of the driving field at twice the frequency $2f$. In case of a unipolar driving field with frequency $f$, $\Delta T_{\mathrm{ex}}(t)$ directly follows $E(t)$ at the same frequency $f$. 
The measurements then allow, e.g., for a comparison of the field dependencies $\Delta T_{\mathrm{ex}}(E)$ for the different driving fields [cf. Figure~\ref{Fig:DifferentDrivingWaveforms}(d)]. For the bulk BZT-12 sample, the field dependencies match very well for the different driving waveforms.

\subsection{\label{Sec:PMN8PT}PMN-8PT thick-film}
In the following, we will show measurements on a $38$~$\mu$m $0.92$Pb(Mg$_{1/3}$Nb$_{2/3}$)O$_3$-$0.08$~PbTiO$_3$ (PMN-8PT) thick film on an inactive alumina substrate as an example for a thick-film application. PMN-PT is one of the best-studied relaxor electrocaloric material systems, where PMN-8PT has attracted special interest due to its operational temperature window around room temperature. The PMN-8PT was synthesized using the columbite route and the sample was screen-printed with bottom and top gold electrodes on an alumina substrate as described in Ref.~\onlinecite{Molin_Ferroelectrics_498_111}. The alumina substrate was directly fixed with a thermally conductive glue on a $2$~mm-thick copper plate attached to the cold-finger of the cryostat. Figure~\ref{Fig:DeltaTvsFrequency}(b) shows the maximum temperature change $\Delta T_{\mathrm{ex}}^{\mathrm{max}}$ for different frequencies of the driving field at a field amplitude of $3.9$~kV/mm and a temperature of $295$~K. The frequency dependence of $\Delta T_{\mathrm{ex}}^{\mathrm{max}}$ shows a qualitatively similar behavior as compared to the bulk BZT-12 sample, with a steep increase of $\Delta T_{\mathrm{ex}}^{\mathrm{max}}$ at low frequencies before reaching an almost frequency independent plateau. In contrast to the bulk BZT-12 sample, $\Delta T_{\mathrm{ex}}^{\mathrm{max}}$, however, steeply increases up to significantly higher frequencies $f > 75$~Hz. This behaviour is expected for the reduced thickness of the PMN-8PT film, which leads to faster thermal equilibration of the sample with the substrate and sample holder. Accordingly, faster modulation is required to reach adiabatic conditions. Figure~\ref{Fig:DeltaTvsFrequency}(b) further compares the frequency dependence of $\Delta T_{\mathrm{ex}}^{\mathrm{max}}$ on an uncoated sample spot and on sample spots coated with a single layer and two layers of black paint with layer thicknesses of approximately $5$~$\mu$m and  $10$~$\mu$m, respectively.\cite{Note2} The frequency dependencies match very well for low frequencies, where the temperature of the layer of paint can follow the temperature change of the sample. While $\Delta T_{\mathrm{ex}}^{\mathrm{max}}$ only shows a slight decrease for the highest frequencies for the single layer of black paint, a strong decrease of $\Delta T_{\mathrm{ex}}^{\mathrm{max}}$ is found for frequencies above approximately $50$~Hz on the sample spot coated with two layers of black paint. In this regime, the thick layer of paint cannot follow the temperature change of the sample anymore, leading to a reduction of the observed temperature change. This example clearly illustrates how advantageous systematic studies of the dynamics of electrocaloric effect are to identify, e.g., extrinsic contributions to the experimentally observed temperature change $\Delta T_{\mathrm{ex}}$. 

The highest temperature change $\Delta T_{\mathrm{ex}}^{\mathrm{max}} \approx 200$~mK for high frequencies is considerably smaller than the adiabatic temperature change observerd in bulk ceramic PMN-8PT samples or PMN-8PT multilayer ceramic samples at comparable electric fields.\cite{Molin_Ferroelectrics_498_111,Molin_JAmCeramSoc_100_2885,Molin_JEurCeramSoc_35_2065,Molin_EnergyTechnol_6_1543} However, we find a relative permittivity $\varepsilon$ with maximum $\varepsilon \approx 12 \times 10^3$ at a temperature of $298$~K and a frequency of $78$~Hz for our samples, which agrees well with literature values for such thick-film samples.\cite{Molin_Ferroelectrics_498_111} These values are strongly reduced as compared to relative permittivities on the order of $25 \times 10^3$ found for bulk ceramics and multilayer ceramics.\cite{Molin_JAmCeramSoc_100_2885,Molin_JEurCeramSoc_35_2065} This strong reduction of the permittivity has been explained by clamping to the substrate and by chemical modification of the PMN-8PT by Li$_2$CO$_3$ which was added as a sintering aid.\cite{Molin_Ferroelectrics_498_111} We therefore expect also a strongly reduced electrocaloric effect in the PMN-8PT thick film, similar to the strongly reduced adiabatic temperature change for doped samples with strongly reduced permittivity.\cite{Molin_JEurCeramSoc_35_2065} In addition, a thick top electrode might also lead to a reduction of the observed temperature change.\cite{Greiner_EnergyTechnol_6_1535}

\subsection{\label{Sec:Polymerfilms}PVDF polymer thin films}
\begin{figure}
	\centering
	\includegraphics[width=1.0\columnwidth]{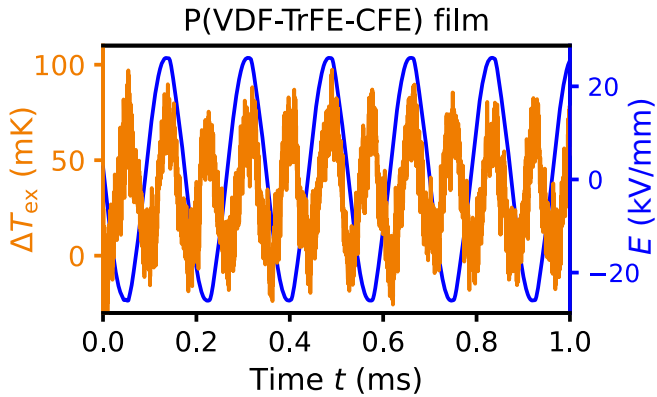}
	\caption{(Color online) Transient of the temperature change $T_{\mathrm{ex}}(t)$ in a 20~$\mu$m thick P(VDF-TrFE-CFE) film for a driving frequency of 5738~Hz. The transient is averaged over 100 measurements.}
	\label{Fig:PolymerTransiente}
\end{figure}

Polymers based on polyvinylidene fluoride (PVDF) are highly interesting candidates for electrocaloric cooling applications\cite{Ma_Science_357_1130} due to their good electrocaloric properties and easy processability. The terpolymers of PVDF with trifluoroethylene-chlorofluoroethylene [P(VDF$_x$-TrFE$_{1-x}$-CFE$_y$)] are relaxor ferroelectric polymers with a large electrocaloric effect near room-temperature.\cite{Neese_Science_321_821,Neese_APL_94_042910} Here, we show measurements on a 20~$\mu$m-thick [P(VDF$_x$-TrFE$_{1-x}$-CFE$_y$)] polymer film as an example for film measurements. The [P(VDF$_x$-TrFE$_{1-x}$-CFE$_y$)] film with $x=0.7$ and $y=0.08$ was prepared by the solution-cast method\cite{Hambal_Polymers_13_1343} and was fixed with a silver conductive epoxy adhesive to a $60~\mu$m-thick Cu foil mounted on the thin-film sample holder described above. Figure~\ref{Fig:PolymerTransiente} shows the average of 100 transients of the temperature change $\Delta T_{\mathrm{ex}}(t)$ for a driving field with amplitude $E_{\mathrm{p}}=25$~kV/mm and frequency of $5738$~Hz as an example for high-frequency measurements as required for adiabatic measurements of film samples on inactive substrates. Notably, the dynamics of the ECE is still well resolved despite the high driving frequency and small driving field amplitude which is considerably lower than in previous studies of the ECE in polymer films.\cite{Li_APL_99_052907,Liu_JAP_126_234102,Jia_APL_103_042903} This example therefore demonstrates the good thermal and temporal resolution of our technique.

\subsection{\label{Sec:Fatigue}Investigation of fatigue in relaxor ferroelectrics}
\begin{figure}
	\centering
	\includegraphics[width=1.0\columnwidth]{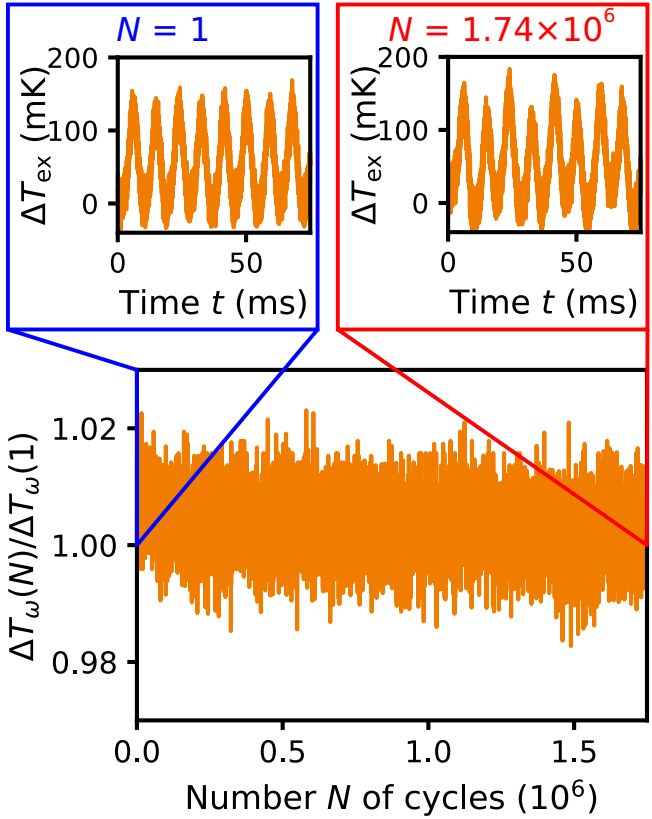}
	\caption{(Color online) Evolution of the first Fourier component $\Delta T_\omega (N)$ of the temperature change for application of $N$ cycles of a unipolar driving electric field with amplitude of $3.9$~kV/mm and frequency of $113$~Hz for a $38$~$\mu$m thick PMN-8PT thick film sample, where  $\Delta T_\omega (N)$ is normalized to the value of  $\Delta T_\omega (1)$ for the first applied electric field cycle. The insets show transients of the temperature change $\Delta T_{\mathrm{ex}}(t)$ at the beginning of the fatigue measurement and after $N=1.74\times10^6$ electric field cycles.}
	\label{Fig:Fatigue}
\end{figure}

Cycling the electric field applied to an electrocaloric material can in general lead to fatigue, i. e. the failure of the material, or to electrical breakdown. Though fatigue is highly important in view of application,\cite{LupascuFatigueFerroelectricCeramics,Lupascu_AdvEngMater_7_882} it has only recently been systematically studied in the context of ECE for large numbers $N > 10^6$ of cycles of the applied electric field.\cite{Fulanovic_JEurCeramSoc_37_5105,Weyland_JEuropCeramSoc_38_551,Bradesko_ActaMaterialia_169_275,delDuca_JAP_128_104102} In the following, we will demonstrate that our technique can also advantageously be applied to the investigation of fatigue in electrocaloric materials. Figure~\ref{Fig:Fatigue} shows a typical measurement protocol for the example of a $38$~$\mu$m-thick PMN-$8$PT thick film. Prior to the fatigue measurement, a transient of the sample temperature change $\Delta T_{\mathrm{ex}}(t)$ is recorded for a unipolar harmonic electric field with frequency $\omega = 2\pi f$.  A Fourier analysis of $\Delta T_{\mathrm{ex}}(t)$ then gives the component $n \omega$, which contributes dominantly. In the example shown in Figure~\ref{Fig:Fatigue}, the first Fourier component at frequency $\omega$ contributes almost 75\% as is expected for unipolar driving. For the measurement of fatigue, this leading harmonic component $\Delta T_{\omega}(N)$ is then measured quasi-continuously by a lock-in amplifier as a function of applied field cycles $N$, with excellent agreement between the Fourier component determined from the full transient and the result of the lock-in amplifier measurement. Figure~\ref{Fig:Fatigue} examplarily shows the evolution of the temperature change $\Delta T_{\omega}(N)$ for an electric field with an amplitude of $3.9$~kV/mm and a frequency of $113$~Hz for more than $1.7\times 10^6$ cycles, where $\Delta T_{\omega}(N)$ is very stable in this case. After a given number of electric field cycles, full transients of the temperature change $\Delta T_{\mathrm{ex}}(t)$ can again be recorded. The example shown in Figure~\ref{Fig:Fatigue} illustrates that this approach can give additional information as the leading Fourier component $\Delta T_{\omega}$ remains constant while the full transient after $N = 1.7\times 10^6$ field cycles shows clear fluctuations of $\Delta T_{\mathrm{ex}}(t)$ on a ms-timescale which were not observed in the beginning. This effect may correlate with the known increasingly dissipative domain system movement due to fatigue.\cite{Nuffer_ActaMaterialia_48_3783,Lupascu_EPL_68_733} More studies are needed for a detailed understanding of the observed changes, which might also reflect the onset of fatigue due to, e.g., vacancy migration\cite{Weyland_JEuropCeramSoc_38_551} or changes in the grain boundary conductivity.\cite{Bradesko_ActaMaterialia_169_275} The example demonstrates, however, the benefits of our technique also for the study of fatigue.

\section{Conclusion}
To summarize, we have introduced a novel direct, contactless method for studying the full dynamics of the electrocaloric effect in ferroelectric samples. The technique is based on a large-bandwidth detection of the infrared radiation emitted by the electrocaloric sample, which allows to investigate transients of the caloric temperature change with mK temperature resolution and sub-ms temporal resolution. The applied electric field and the induced polarization are recorded simultaneously with the temperature change, thus allowing to additionally also obtain the full electric-field dependencies $\Delta T(E)$ of the temperature change and $P(E)$ of the polarization within one single measurement. We expect that the possibility to correlate the dynamics of the electrocaloric effect with the polarization dynamics will allow for a deeper understanding of the electrocaloric effect as, on the one hand, the influence of, e.g. the relaxational behavior of the polarization will become accessible, and, on the other hand, spurious effects like heat exchange with the sourrounding can be identified. In addition, the polarization dependence of the adiabatic temperature change $\Delta T(P)$ becomes accessible, which can be compared to predictions by, e.g., phenomenological Landau-Devonshire theory.\cite{FatuzzoMerzFerroelectricity} The combined measurement of electrocaloric and polarization properties will also be beneficial for the study of materials exhibiting strong ageing, where simultaneous measurements guarantee consistent data sets.

We have further demonstrated high frequency measurements of the electrocaloric effect in polymer films for frequencies exceeding several kHz. As a perspective, the technique can be scaled up to even higher measurement frequencies, thus allowing for the investigation of thin films of electrocaloric materials as required for the reliable characterization of novel electrocaloric materials synthesized as thin films on substrates. Quasi-continuous measurements of the temperature change $\Delta T$ in addition allow for the investigation of fatigue for large numbers of applied electric field cycles. 
\section*{Acknowledgments}
This work is funded by the Deutsche Forschungsgemeinschaft (DFG, German
Research Foundation) - 418847609.

\section*{Author declarations}
\subsection*{Conflict of Interest}
The authors have no conflicts to disclose.

\subsection*{Author Contributions}
\textbf{J. Fischer:} Data curation (lead); Formal analysis (lead); Investigation (equal); Writing - review \& editing (equal). \textbf{J. D\"{o}ntgen:} Formal analysis (supporting); Investigation (supporting); Writing - review \& editing (equal). \textbf{C. Molin:} Resources (equal);  Writing - review \& editing (equal). \textbf{S. E. Gebhardt:} Resources (equal);  Writing - review \& editing (equal). \textbf{Y. Hambal:} Resources (equal). \textbf{V. V. Shvartsman:} Resources (equal);  Writing - review \& editing (equal). \textbf{D. C. Lupascu:} Resources (equal);  Writing - review \& editing (equal). \textbf{D. H\"{a}gele:} Conceptualization (supporting); Formal analysis (supporting); Investigation (supporting); Supervision (equal); Writing - review \& editing (equal). \textbf{J. Rudolph:} Conceptualization (lead); Formal analysis (supporting); Investigation (equal); Supervision (equal); Writing – original draft (lead); Writing - review \& editing (lead).

\section*{Data availability}
The data that support the findings of this study are available from the corresponding author upon reasonable request.

\section*{References}

\begin{thebibliography}{61}%
	\makeatletter
	\providecommand \@ifxundefined [1]{%
		\@ifx{#1\undefined}
	}%
	\providecommand \@ifnum [1]{%
		\ifnum #1\expandafter \@firstoftwo
		\else \expandafter \@secondoftwo
		\fi
	}%
	\providecommand \@ifx [1]{%
		\ifx #1\expandafter \@firstoftwo
		\else \expandafter \@secondoftwo
		\fi
	}%
	\providecommand \natexlab [1]{#1}%
	\providecommand \enquote  [1]{``#1''}%
	\providecommand \bibnamefont  [1]{#1}%
	\providecommand \bibfnamefont [1]{#1}%
	\providecommand \citenamefont [1]{#1}%
	\providecommand \href@noop [0]{\@secondoftwo}%
	\providecommand \href [0]{\begingroup \@sanitize@url \@href}%
	\providecommand \@href[1]{\@@startlink{#1}\@@href}%
	\providecommand \@@href[1]{\endgroup#1\@@endlink}%
	\providecommand \@sanitize@url [0]{\catcode `\\12\catcode `\$12\catcode
		`\&12\catcode `\#12\catcode `\^12\catcode `\_12\catcode `\%12\relax}%
	\providecommand \@@startlink[1]{}%
	\providecommand \@@endlink[0]{}%
	\providecommand \url  [0]{\begingroup\@sanitize@url \@url }%
	\providecommand \@url [1]{\endgroup\@href {#1}{\urlprefix }}%
	\providecommand \urlprefix  [0]{URL }%
	\providecommand \Eprint [0]{\href }%
	\providecommand \doibase [0]{https://doi.org/}%
	\providecommand \selectlanguage [0]{\@gobble}%
	\providecommand \bibinfo  [0]{\@secondoftwo}%
	\providecommand \bibfield  [0]{\@secondoftwo}%
	\providecommand \translation [1]{[#1]}%
	\providecommand \BibitemOpen [0]{}%
	\providecommand \bibitemStop [0]{}%
	\providecommand \bibitemNoStop [0]{.\EOS\space}%
	\providecommand \EOS [0]{\spacefactor3000\relax}%
	\providecommand \BibitemShut  [1]{\csname bibitem#1\endcsname}%
	\let\auto@bib@innerbib\@empty
	\bibitem [{\citenamefont {Kitanovski}\ \emph {et~al.}(2015)\citenamefont
		{Kitanovski}, \citenamefont {Plaznik}, \citenamefont {Tomc},\ and\
		\citenamefont {Poredo\v{s}}}]{Kitanovski_IntJRefrig_57_288}%
	\BibitemOpen
	\bibfield  {author} {\bibinfo {author} {\bibfnamefont {A.}~\bibnamefont
			{Kitanovski}}, \bibinfo {author} {\bibfnamefont {U.}~\bibnamefont {Plaznik}},
		\bibinfo {author} {\bibfnamefont {U.}~\bibnamefont {Tomc}},\ and\ \bibinfo
		{author} {\bibfnamefont {A.}~\bibnamefont {Poredo\v{s}}},\ }\bibfield
	{title} {\enquote {\bibinfo {title} {Present and future caloric refrigeration
				and heatpump technologies},}\ }\href@noop {} {\bibfield  {journal} {\bibinfo
			{journal} {Int. J. Refrig.}\ }\textbf {\bibinfo {volume} {57}},\ \bibinfo
		{pages} {288--298} (\bibinfo {year} {2015})}\BibitemShut {NoStop}%
	\bibitem [{\citenamefont {F\"{a}hler}\ \emph {et~al.}(2012)\citenamefont
		{F\"{a}hler}, \citenamefont {R\"{o}{\ss}ler}, \citenamefont {Kastner},
		\citenamefont {Eckert}, \citenamefont {Eggeler}, \citenamefont {Emmerich},
		\citenamefont {Entel}, \citenamefont {M\"{u}ller}, \citenamefont {Quandt},\
		and\ \citenamefont {Albe}}]{Faehler_AdvEngMater_14_10}%
	\BibitemOpen
	\bibfield  {author} {\bibinfo {author} {\bibfnamefont {S.}~\bibnamefont
			{F\"{a}hler}}, \bibinfo {author} {\bibfnamefont {U.~K.}\ \bibnamefont
			{R\"{o}{\ss}ler}}, \bibinfo {author} {\bibfnamefont {O.}~\bibnamefont
			{Kastner}}, \bibinfo {author} {\bibfnamefont {J.}~\bibnamefont {Eckert}},
		\bibinfo {author} {\bibfnamefont {G.}~\bibnamefont {Eggeler}}, \bibinfo
		{author} {\bibfnamefont {H.}~\bibnamefont {Emmerich}}, \bibinfo {author}
		{\bibfnamefont {P.}~\bibnamefont {Entel}}, \bibinfo {author} {\bibfnamefont
			{S.}~\bibnamefont {M\"{u}ller}}, \bibinfo {author} {\bibfnamefont
			{E.}~\bibnamefont {Quandt}},\ and\ \bibinfo {author} {\bibfnamefont
			{K.}~\bibnamefont {Albe}},\ }\bibfield  {title} {\enquote {\bibinfo {title}
			{Caloric effects in ferroic materials: New concepts for cooling},}\
	}\href@noop {} {\bibfield  {journal} {\bibinfo  {journal} {Adv. Eng. Mater.}\
		}\textbf {\bibinfo {volume} {14}},\ \bibinfo {pages} {10--19} (\bibinfo
		{year} {2012})}\BibitemShut {NoStop}%
	\bibitem [{\citenamefont {Moya}\ and\ \citenamefont
		{Mathur}(2020)}]{Moya_Science_370_797}%
	\BibitemOpen
	\bibfield  {author} {\bibinfo {author} {\bibfnamefont {X.}~\bibnamefont
			{Moya}}\ and\ \bibinfo {author} {\bibfnamefont {N.~D.}\ \bibnamefont
			{Mathur}},\ }\bibfield  {title} {\enquote {\bibinfo {title} {{Caloric
					materials for cooling and heating}},}\ }\href@noop {} {\bibfield  {journal}
		{\bibinfo  {journal} {Science}\ }\textbf {\bibinfo {volume} {370}},\ \bibinfo
		{pages} {797--803} (\bibinfo {year} {2020})}\BibitemShut {NoStop}%
	\bibitem [{\citenamefont {Moya}, \citenamefont {Kar-Narayan},\ and\
		\citenamefont {Mathur}(2014)}]{Moya_NatureMater_13_439}%
	\BibitemOpen
	\bibfield  {author} {\bibinfo {author} {\bibfnamefont {X.}~\bibnamefont
			{Moya}}, \bibinfo {author} {\bibfnamefont {S.}~\bibnamefont {Kar-Narayan}},\
		and\ \bibinfo {author} {\bibfnamefont {N.~D.}\ \bibnamefont {Mathur}},\
	}\bibfield  {title} {\enquote {\bibinfo {title} {{Caloric materials near
					ferroic phase transitions}},}\ }\href@noop {} {\bibfield  {journal} {\bibinfo
			{journal} {Nature Mater.}\ }\textbf {\bibinfo {volume} {13}},\ \bibinfo
		{pages} {439--450} (\bibinfo {year} {2014})}\BibitemShut {NoStop}%
	\bibitem [{\citenamefont {Kobeko}\ and\ \citenamefont
		{Kurtschatov}(1930)}]{Kobeko_ZPhys_66_192}%
	\BibitemOpen
	\bibfield  {author} {\bibinfo {author} {\bibfnamefont {P.}~\bibnamefont
			{Kobeko}}\ and\ \bibinfo {author} {\bibfnamefont {J.}~\bibnamefont
			{Kurtschatov}},\ }\bibfield  {title} {\enquote {\bibinfo {title}
			{{Dielektrische Eigenschaften der Seignettesalzkristalle}},}\ }\href@noop {}
	{\bibfield  {journal} {\bibinfo  {journal} {Z. Physik}\ }\textbf {\bibinfo
			{volume} {66}},\ \bibinfo {pages} {192--205} (\bibinfo {year}
		{1930})}\BibitemShut {NoStop}%
	\bibitem [{\citenamefont {Mischenko}\ \emph {et~al.}(2006)\citenamefont
		{Mischenko}, \citenamefont {Zhang}, \citenamefont {Scott}, \citenamefont
		{Whatmore},\ and\ \citenamefont {Mathur}}]{Mischenko_Science_311_1270}%
	\BibitemOpen
	\bibfield  {author} {\bibinfo {author} {\bibfnamefont {A.~S.}\ \bibnamefont
			{Mischenko}}, \bibinfo {author} {\bibfnamefont {Q.}~\bibnamefont {Zhang}},
		\bibinfo {author} {\bibfnamefont {J.~F.}\ \bibnamefont {Scott}}, \bibinfo
		{author} {\bibfnamefont {R.~W.}\ \bibnamefont {Whatmore}},\ and\ \bibinfo
		{author} {\bibfnamefont {N.~D.}\ \bibnamefont {Mathur}},\ }\bibfield  {title}
	{\enquote {\bibinfo {title} {{Giant Electrocaloric Effect in Thin-Film
					PbZr$_{0.95}$Ti$_{0.05}$O$_{3}$}},}\ }\href@noop {} {\bibfield  {journal}
		{\bibinfo  {journal} {Science}\ }\textbf {\bibinfo {volume} {311}},\ \bibinfo
		{pages} {1270--1271} (\bibinfo {year} {2006})}\BibitemShut {NoStop}%
	\bibitem [{\citenamefont {Neese}\ \emph {et~al.}(2008)\citenamefont {Neese},
		\citenamefont {Chu}, \citenamefont {Lu}, \citenamefont {Wang}, \citenamefont
		{Furman},\ and\ \citenamefont {Zhang}}]{Neese_Science_321_821}%
	\BibitemOpen
	\bibfield  {author} {\bibinfo {author} {\bibfnamefont {B.}~\bibnamefont
			{Neese}}, \bibinfo {author} {\bibfnamefont {B.}~\bibnamefont {Chu}}, \bibinfo
		{author} {\bibfnamefont {S.-G.}\ \bibnamefont {Lu}}, \bibinfo {author}
		{\bibfnamefont {Y.}~\bibnamefont {Wang}}, \bibinfo {author} {\bibfnamefont
			{E.}~\bibnamefont {Furman}},\ and\ \bibinfo {author} {\bibfnamefont {Q.~M.}\
			\bibnamefont {Zhang}},\ }\bibfield  {title} {\enquote {\bibinfo {title}
			{{Large Electrocaloric Effect in Ferroelectric Polymers Near Room
					Temperature}},}\ }\href@noop {} {\bibfield  {journal} {\bibinfo  {journal}
			{Science}\ }\textbf {\bibinfo {volume} {321}},\ \bibinfo {pages} {821--823}
		(\bibinfo {year} {2008})}\BibitemShut {NoStop}%
	\bibitem [{\citenamefont {Torell\'{o}}\ and\ \citenamefont
		{Defay}(2022)}]{Torello_AdvElectronMater_8_2101031}%
	\BibitemOpen
	\bibfield  {author} {\bibinfo {author} {\bibfnamefont {A.}~\bibnamefont
			{Torell\'{o}}}\ and\ \bibinfo {author} {\bibfnamefont {E.}~\bibnamefont
			{Defay}},\ }\bibfield  {title} {\enquote {\bibinfo {title} {{Electrocaloric
					Coolers: A Review}},}\ }\href@noop {} {\bibfield  {journal} {\bibinfo
			{journal} {Adv. Electron. Mater.}\ }\textbf {\bibinfo {volume} {8}},\
		\bibinfo {pages} {2101031} (\bibinfo {year} {2022})}\BibitemShut {NoStop}%
	\bibitem [{\citenamefont {Liu}, \citenamefont {Scott},\ and\ \citenamefont
		{Dkhil}(2016)}]{Liu_ApplPhysRev_3_0311022}%
	\BibitemOpen
	\bibfield  {author} {\bibinfo {author} {\bibfnamefont {Y.}~\bibnamefont
			{Liu}}, \bibinfo {author} {\bibfnamefont {J.~F.}\ \bibnamefont {Scott}},\
		and\ \bibinfo {author} {\bibfnamefont {B.}~\bibnamefont {Dkhil}},\ }\bibfield
	{title} {\enquote {\bibinfo {title} {Direct and indirect measurements on
				electrocaloric effect: Recent developments and perspectives},}\ }\href@noop
	{} {\bibfield  {journal} {\bibinfo  {journal} {Appl. Phys. Rev.}\ }\textbf
		{\bibinfo {volume} {3}},\ \bibinfo {pages} {031102} (\bibinfo {year}
		{2016})}\BibitemShut {NoStop}%
	\bibitem [{\citenamefont {Chen}\ \emph {et~al.}(2021)\citenamefont {Chen},
		\citenamefont {Li}, \citenamefont {Jian}, \citenamefont {Hambal},
		\citenamefont {Lu}, \citenamefont {Shvartsman}, \citenamefont {Lupascu},\
		and\ \citenamefont {Zhang}}]{Chen_APL_118_122904}%
	\BibitemOpen
	\bibfield  {author} {\bibinfo {author} {\bibfnamefont {X.}~\bibnamefont
			{Chen}}, \bibinfo {author} {\bibfnamefont {S.}~\bibnamefont {Li}}, \bibinfo
		{author} {\bibfnamefont {X.}~\bibnamefont {Jian}}, \bibinfo {author}
		{\bibfnamefont {Y.}~\bibnamefont {Hambal}}, \bibinfo {author} {\bibfnamefont
			{S.-G.}\ \bibnamefont {Lu}}, \bibinfo {author} {\bibfnamefont {V.~V.}\
			\bibnamefont {Shvartsman}}, \bibinfo {author} {\bibfnamefont {D.~C.}\
			\bibnamefont {Lupascu}},\ and\ \bibinfo {author} {\bibfnamefont {Q.~M.}\
			\bibnamefont {Zhang}},\ }\bibfield  {title} {\enquote {\bibinfo {title}
			{Maxwell relation, giant (negative) electrocaloric effect, and polarization
				hysteresis},}\ }\href@noop {} {\bibfield  {journal} {\bibinfo  {journal}
			{Appl. Phys. Lett.}\ }\textbf {\bibinfo {volume} {118}},\ \bibinfo {pages}
		{122904} (\bibinfo {year} {2021})}\BibitemShut {NoStop}%
	\bibitem [{\citenamefont {Cheng}\ \emph {et~al.}(2019)\citenamefont {Cheng},
		\citenamefont {Weyland}, \citenamefont {Novak},\ and\ \citenamefont
		{Li}}]{Cheng_PhysStatusSolidiA_216_1900684}%
	\BibitemOpen
	\bibfield  {author} {\bibinfo {author} {\bibfnamefont {X.}~\bibnamefont
			{Cheng}}, \bibinfo {author} {\bibfnamefont {F.}~\bibnamefont {Weyland}},
		\bibinfo {author} {\bibfnamefont {N.}~\bibnamefont {Novak}},\ and\ \bibinfo
		{author} {\bibfnamefont {Y.}~\bibnamefont {Li}},\ }\bibfield  {title}
	{\enquote {\bibinfo {title} {Indirect electrocaloric evaluation: Influence of
				polarization hysteresis measurement frequency},}\ }\href@noop {} {\bibfield
		{journal} {\bibinfo  {journal} {Phys. Status Solidi A}\ }\textbf {\bibinfo
			{volume} {216}},\ \bibinfo {pages} {1900684} (\bibinfo {year}
		{2019})}\BibitemShut {NoStop}%
	\bibitem [{\citenamefont {Kutnjak}\ and\ \citenamefont
		{Ro\v{z}i\v{c}}(2014)}]{Kutnjak_inElectrocaloricMaterials}%
	\BibitemOpen
	\bibfield  {author} {\bibinfo {author} {\bibfnamefont {Z.}~\bibnamefont
			{Kutnjak}}\ and\ \bibinfo {author} {\bibfnamefont {B.}~\bibnamefont
			{Ro\v{z}i\v{c}}},\ }\enquote {\bibinfo {title} {{Indirect and Direct
				Measurements of the Electrocaloric Effect}},}\ in\ \href@noop {} {\emph
		{\bibinfo {booktitle} {{Electrocaloric Materials: New Generation of
					Coolers}}}},\ \bibinfo {series} {Engineering Materials}, Vol.~\bibinfo
	{volume} {34},\ \bibinfo {editor} {edited by\ \bibinfo {editor}
		{\bibfnamefont {T.}~\bibnamefont {Correia}}\ and\ \bibinfo {editor}
		{\bibfnamefont {Q.}~\bibnamefont {Zhang}}}\ (\bibinfo  {publisher}
	{Springer},\ \bibinfo {year} {2014})\ Chap.~\bibinfo {chapter}
	{7}\BibitemShut {NoStop}%
	\bibitem [{\citenamefont {Sanlialp}\ \emph {et~al.}(2015)\citenamefont
		{Sanlialp}, \citenamefont {Shvartsman}, \citenamefont {Acosta}, \citenamefont
		{Dkhil},\ and\ \citenamefont {Lupascu}}]{Sanlialp_APL_106_062901}%
	\BibitemOpen
	\bibfield  {author} {\bibinfo {author} {\bibfnamefont {M.}~\bibnamefont
			{Sanlialp}}, \bibinfo {author} {\bibfnamefont {V.~V.}\ \bibnamefont
			{Shvartsman}}, \bibinfo {author} {\bibfnamefont {M.}~\bibnamefont {Acosta}},
		\bibinfo {author} {\bibfnamefont {B.}~\bibnamefont {Dkhil}},\ and\ \bibinfo
		{author} {\bibfnamefont {D.~C.}\ \bibnamefont {Lupascu}},\ }\bibfield
	{title} {\enquote {\bibinfo {title} {{Strong electrocaloric effect in
					lead-free
					0.65{Ba}(Zr$_{0.2}$Ti$_{0.8}$)O$_{3}$-0.35(Ba$_{0.7}$Ca$_{0.3}$)TiO$_{3}$
					ceramics obtained by direct measurements}},}\ }\href@noop {} {\bibfield
		{journal} {\bibinfo  {journal} {Appl. Phys. Lett.}\ }\textbf {\bibinfo
			{volume} {106}},\ \bibinfo {pages} {062901} (\bibinfo {year}
		{2015})}\BibitemShut {NoStop}%
	\bibitem [{\citenamefont {Sanlialp}\ \emph {et~al.}(2017)\citenamefont
		{Sanlialp}, \citenamefont {Luo}, \citenamefont {Shvartsman}, \citenamefont
		{Wei}, \citenamefont {Liu}, \citenamefont {Dkhil},\ and\ \citenamefont
		{Lupascu}}]{Sanlialp_APL_111_173903}%
	\BibitemOpen
	\bibfield  {author} {\bibinfo {author} {\bibfnamefont {M.}~\bibnamefont
			{Sanlialp}}, \bibinfo {author} {\bibfnamefont {Z.}~\bibnamefont {Luo}},
		\bibinfo {author} {\bibfnamefont {V.~V.}\ \bibnamefont {Shvartsman}},
		\bibinfo {author} {\bibfnamefont {X.}~\bibnamefont {Wei}}, \bibinfo {author}
		{\bibfnamefont {Y.}~\bibnamefont {Liu}}, \bibinfo {author} {\bibfnamefont
			{B.}~\bibnamefont {Dkhil}},\ and\ \bibinfo {author} {\bibfnamefont {D.~C.}\
			\bibnamefont {Lupascu}},\ }\bibfield  {title} {\enquote {\bibinfo {title}
			{{Direct measurement of electrocaloric effect in lead-free
					Ba(Sn$_x$Ti$_{1-x}$)O$_3 $ ceramics}},}\ }\href@noop {} {\bibfield  {journal}
		{\bibinfo  {journal} {Appl. Phys. Lett.}\ }\textbf {\bibinfo {volume}
			{111}},\ \bibinfo {pages} {173903} (\bibinfo {year} {2017})}\BibitemShut
	{NoStop}%
	\bibitem [{\citenamefont {Le~Goupil}\ \emph {et~al.}(2015)\citenamefont
		{Le~Goupil}, \citenamefont {Bennett}, \citenamefont {Axelsson}, \citenamefont
		{Valant}, \citenamefont {Berenov}, \citenamefont {Bell}, \citenamefont
		{Comyn},\ and\ \citenamefont {Alford}}]{LeGoupil_APL_107_172903}%
	\BibitemOpen
	\bibfield  {author} {\bibinfo {author} {\bibfnamefont {F.}~\bibnamefont
			{Le~Goupil}}, \bibinfo {author} {\bibfnamefont {J.}~\bibnamefont {Bennett}},
		\bibinfo {author} {\bibfnamefont {A.-K.}\ \bibnamefont {Axelsson}}, \bibinfo
		{author} {\bibfnamefont {M.}~\bibnamefont {Valant}}, \bibinfo {author}
		{\bibfnamefont {A.}~\bibnamefont {Berenov}}, \bibinfo {author} {\bibfnamefont
			{A.~J.}\ \bibnamefont {Bell}}, \bibinfo {author} {\bibfnamefont {T.~P.}\
			\bibnamefont {Comyn}},\ and\ \bibinfo {author} {\bibfnamefont {N.~M.}\
			\bibnamefont {Alford}},\ }\bibfield  {title} {\enquote {\bibinfo {title}
			{{Electrocaloric enhancement near the morphotropic phase boundary in
					lead-free NBT-KBT ceramics}},}\ }\href@noop {} {\bibfield  {journal}
		{\bibinfo  {journal} {Appl. Phys. Lett.}\ }\textbf {\bibinfo {volume}
			{107}},\ \bibinfo {pages} {172903} (\bibinfo {year} {2015})}\BibitemShut
	{NoStop}%
	\bibitem [{\citenamefont {Lu}\ \emph {et~al.}(2010{\natexlab{a}})\citenamefont
		{Lu}, \citenamefont {Rozic}, \citenamefont {Zhang}, \citenamefont {Kutnjak},
		\citenamefont {Pirc}, \citenamefont {Lin}, \citenamefont {Li},\ and\
		\citenamefont {Gorny}}]{Lu_APL_97_202901}%
	\BibitemOpen
	\bibfield  {author} {\bibinfo {author} {\bibfnamefont {S.~G.}\ \bibnamefont
			{Lu}}, \bibinfo {author} {\bibfnamefont {B.}~\bibnamefont {Rozic}}, \bibinfo
		{author} {\bibfnamefont {Q.~M.}\ \bibnamefont {Zhang}}, \bibinfo {author}
		{\bibfnamefont {Z.}~\bibnamefont {Kutnjak}}, \bibinfo {author} {\bibfnamefont
			{R.}~\bibnamefont {Pirc}}, \bibinfo {author} {\bibfnamefont {M.}~\bibnamefont
			{Lin}}, \bibinfo {author} {\bibfnamefont {X.}~\bibnamefont {Li}},\ and\
		\bibinfo {author} {\bibfnamefont {L.}~\bibnamefont {Gorny}},\ }\bibfield
	{title} {\enquote {\bibinfo {title} {{Comparison of directly and indirectly
					measured electrocaloric effect in relaxor ferroelectric polymers}},}\
	}\href@noop {} {\bibfield  {journal} {\bibinfo  {journal} {Appl. Phys.
				Lett.}\ }\textbf {\bibinfo {volume} {97}},\ \bibinfo {pages} {202901}
		(\bibinfo {year} {2010}{\natexlab{a}})}\BibitemShut {NoStop}%
	\bibitem [{\citenamefont {Birks}\ \emph {et~al.}(2017)\citenamefont {Birks},
		\citenamefont {Dunce}, \citenamefont {Peräntie}, \citenamefont {Hagberg},\
		and\ \citenamefont {Sternberg}}]{Birks_JAP_121_224102}%
	\BibitemOpen
	\bibfield  {author} {\bibinfo {author} {\bibfnamefont {E.}~\bibnamefont
			{Birks}}, \bibinfo {author} {\bibfnamefont {M.}~\bibnamefont {Dunce}},
		\bibinfo {author} {\bibfnamefont {J.}~\bibnamefont {Peräntie}}, \bibinfo
		{author} {\bibfnamefont {J.}~\bibnamefont {Hagberg}},\ and\ \bibinfo {author}
		{\bibfnamefont {A.}~\bibnamefont {Sternberg}},\ }\bibfield  {title} {\enquote
		{\bibinfo {title} {{Direct and indirect determination of electrocaloric
					effect in Na$_{0.5}$Bi$_{0.5}$TiO$_3$}},}\ }\href@noop {} {\bibfield
		{journal} {\bibinfo  {journal} {J. Appl. Phys.}\ }\textbf {\bibinfo {volume}
			{121}},\ \bibinfo {pages} {224102} (\bibinfo {year} {2017})}\BibitemShut
	{NoStop}%
	\bibitem [{\citenamefont {Pandya}\ \emph {et~al.}(2017)\citenamefont {Pandya},
		\citenamefont {Wilbur}, \citenamefont {Bhatia}, \citenamefont {Damodaran},
		\citenamefont {Monachon}, \citenamefont {Dasgupta}, \citenamefont {King},
		\citenamefont {Dames},\ and\ \citenamefont
		{Martin}}]{Pandya_PRAppl_7_034025}%
	\BibitemOpen
	\bibfield  {author} {\bibinfo {author} {\bibfnamefont {S.}~\bibnamefont
			{Pandya}}, \bibinfo {author} {\bibfnamefont {J.~D.}\ \bibnamefont {Wilbur}},
		\bibinfo {author} {\bibfnamefont {B.}~\bibnamefont {Bhatia}}, \bibinfo
		{author} {\bibfnamefont {A.~R.}\ \bibnamefont {Damodaran}}, \bibinfo {author}
		{\bibfnamefont {C.}~\bibnamefont {Monachon}}, \bibinfo {author}
		{\bibfnamefont {A.}~\bibnamefont {Dasgupta}}, \bibinfo {author}
		{\bibfnamefont {W.~P.}\ \bibnamefont {King}}, \bibinfo {author}
		{\bibfnamefont {C.}~\bibnamefont {Dames}},\ and\ \bibinfo {author}
		{\bibfnamefont {L.~W.}\ \bibnamefont {Martin}},\ }\bibfield  {title}
	{\enquote {\bibinfo {title} {Direct measurement of pyroelectric and
				electrocaloric effects in thin films},}\ }\href@noop {} {\bibfield  {journal}
		{\bibinfo  {journal} {Phys. Rev. Appl.}\ }\textbf {\bibinfo {volume} {7}},\
		\bibinfo {pages} {034025} (\bibinfo {year} {2017})}\BibitemShut {NoStop}%
	\bibitem [{\citenamefont {Matsushita}\ \emph {et~al.}(2020)\citenamefont
		{Matsushita}, \citenamefont {Yoshimura}, \citenamefont {Kiriya},\ and\
		\citenamefont {Fujimura}}]{Matsushita_ApplPhysExpr_13_041007}%
	\BibitemOpen
	\bibfield  {author} {\bibinfo {author} {\bibfnamefont {Y.}~\bibnamefont
			{Matsushita}}, \bibinfo {author} {\bibfnamefont {T.}~\bibnamefont
			{Yoshimura}}, \bibinfo {author} {\bibfnamefont {D.}~\bibnamefont {Kiriya}},\
		and\ \bibinfo {author} {\bibfnamefont {N.}~\bibnamefont {Fujimura}},\
	}\bibfield  {title} {\enquote {\bibinfo {title} {Investigation of the
				electrocaloric effect in ferroelectric polymer film through direct
				measurement under alternating electric field},}\ }\href@noop {} {\bibfield
		{journal} {\bibinfo  {journal} {Appl. Phys. Expr.}\ }\textbf {\bibinfo
			{volume} {13}},\ \bibinfo {pages} {041007} (\bibinfo {year}
		{2020})}\BibitemShut {NoStop}%
	\bibitem [{\citenamefont {Matsushita}\ \emph {et~al.}(2016)\citenamefont
		{Matsushita}, \citenamefont {Nochida}, \citenamefont {Yoshimura},\ and\
		\citenamefont {Fujimura}}]{Matsushita_JpnJApplPhys_55_10TB04}%
	\BibitemOpen
	\bibfield  {author} {\bibinfo {author} {\bibfnamefont {Y.}~\bibnamefont
			{Matsushita}}, \bibinfo {author} {\bibfnamefont {A.}~\bibnamefont {Nochida}},
		\bibinfo {author} {\bibfnamefont {T.}~\bibnamefont {Yoshimura}},\ and\
		\bibinfo {author} {\bibfnamefont {N.}~\bibnamefont {Fujimura}},\ }\bibfield
	{title} {\enquote {\bibinfo {title} {Direct measurements of electrocaloric
				effect in ferroelectrics using thin--film thermocouples},}\ }\href@noop {}
	{\bibfield  {journal} {\bibinfo  {journal} {Jpn. J. Appl. Phys.}\ }\textbf
		{\bibinfo {volume} {55}},\ \bibinfo {pages} {10TB04} (\bibinfo {year}
		{2016})}\BibitemShut {NoStop}%
	\bibitem [{\citenamefont {Lu}\ \emph {et~al.}(2010{\natexlab{b}})\citenamefont
		{Lu}, \citenamefont {Rozic}, \citenamefont {Zhang}, \citenamefont {Kutnjak},
		\citenamefont {Li}, \citenamefont {Furman}, \citenamefont {Gorny},
		\citenamefont {Lin}, \citenamefont {Malic}, \citenamefont {Kosec},
		\citenamefont {Blinc},\ and\ \citenamefont {Pirc}}]{Lu_APL_97_162904}%
	\BibitemOpen
	\bibfield  {author} {\bibinfo {author} {\bibfnamefont {S.~G.}\ \bibnamefont
			{Lu}}, \bibinfo {author} {\bibfnamefont {B.}~\bibnamefont {Rozic}}, \bibinfo
		{author} {\bibfnamefont {Q.~M.}\ \bibnamefont {Zhang}}, \bibinfo {author}
		{\bibfnamefont {Z.}~\bibnamefont {Kutnjak}}, \bibinfo {author} {\bibfnamefont
			{X.}~\bibnamefont {Li}}, \bibinfo {author} {\bibfnamefont {E.}~\bibnamefont
			{Furman}}, \bibinfo {author} {\bibfnamefont {L.~J.}\ \bibnamefont {Gorny}},
		\bibinfo {author} {\bibfnamefont {M.}~\bibnamefont {Lin}}, \bibinfo {author}
		{\bibfnamefont {B.}~\bibnamefont {Malic}}, \bibinfo {author} {\bibfnamefont
			{M.}~\bibnamefont {Kosec}}, \bibinfo {author} {\bibfnamefont
			{R.}~\bibnamefont {Blinc}},\ and\ \bibinfo {author} {\bibfnamefont
			{R.}~\bibnamefont {Pirc}},\ }\bibfield  {title} {\enquote {\bibinfo {title}
			{Organic and inorganic relaxor ferroelectrics with giant electrocaloric
				effect},}\ }\href@noop {} {\bibfield  {journal} {\bibinfo  {journal} {Appl.
				Phys. Lett.}\ }\textbf {\bibinfo {volume} {97}},\ \bibinfo {pages} {162904}
		(\bibinfo {year} {2010}{\natexlab{b}})}\BibitemShut {NoStop}%
	\bibitem [{\citenamefont {Sotnikova}\ \emph {et~al.}(2020)\citenamefont
		{Sotnikova}, \citenamefont {Gavrilov}, \citenamefont {Kapralov},
		\citenamefont {Muratikov},\ and\ \citenamefont
		{Smirnova}}]{Sotnikova_RSI_91_015119}%
	\BibitemOpen
	\bibfield  {author} {\bibinfo {author} {\bibfnamefont {G.~Y.}\ \bibnamefont
			{Sotnikova}}, \bibinfo {author} {\bibfnamefont {G.~A.}\ \bibnamefont
			{Gavrilov}}, \bibinfo {author} {\bibfnamefont {A.~A.}\ \bibnamefont
			{Kapralov}}, \bibinfo {author} {\bibfnamefont {K.~L.}\ \bibnamefont
			{Muratikov}},\ and\ \bibinfo {author} {\bibfnamefont {E.~P.}\ \bibnamefont
			{Smirnova}},\ }\bibfield  {title} {\enquote {\bibinfo {title} {Mid-infrared
				radiation technique for direct pyroelectric and electrocaloric
				measurements},}\ }\href@noop {} {\bibfield  {journal} {\bibinfo  {journal}
			{Rev. Sci. Inst.}\ }\textbf {\bibinfo {volume} {91}},\ \bibinfo {pages}
		{015119} (\bibinfo {year} {2020})}\BibitemShut {NoStop}%
	\bibitem [{\citenamefont {Kar-Narayan}\ \emph {et~al.}(2013)\citenamefont
		{Kar-Narayan}, \citenamefont {Crossley}, \citenamefont {Moya}, \citenamefont
		{Kovacova}, \citenamefont {Abergel}, \citenamefont {Bontempi}, \citenamefont
		{Baier}, \citenamefont {Defay},\ and\ \citenamefont
		{Mathur}}]{KarNarayan_APL_102_032903}%
	\BibitemOpen
	\bibfield  {author} {\bibinfo {author} {\bibfnamefont {S.}~\bibnamefont
			{Kar-Narayan}}, \bibinfo {author} {\bibfnamefont {S.}~\bibnamefont
			{Crossley}}, \bibinfo {author} {\bibfnamefont {X.}~\bibnamefont {Moya}},
		\bibinfo {author} {\bibfnamefont {V.}~\bibnamefont {Kovacova}}, \bibinfo
		{author} {\bibfnamefont {J.}~\bibnamefont {Abergel}}, \bibinfo {author}
		{\bibfnamefont {A.}~\bibnamefont {Bontempi}}, \bibinfo {author}
		{\bibfnamefont {N.}~\bibnamefont {Baier}}, \bibinfo {author} {\bibfnamefont
			{E.}~\bibnamefont {Defay}},\ and\ \bibinfo {author} {\bibfnamefont {N.~D.}\
			\bibnamefont {Mathur}},\ }\bibfield  {title} {\enquote {\bibinfo {title}
			{Direct electrocaloric measurements of a multilayer capacitor using scanning
				thermal microscopy and infra-red imaging},}\ }\href@noop {} {\bibfield
		{journal} {\bibinfo  {journal} {Appl. Phys. Lett.}\ }\textbf {\bibinfo
			{volume} {102}},\ \bibinfo {pages} {032903} (\bibinfo {year}
		{2013})}\BibitemShut {NoStop}%
	\bibitem [{\citenamefont {Sebald}\ \emph {et~al.}(2012)\citenamefont {Sebald},
		\citenamefont {Seveyrat}, \citenamefont {Capsal}, \citenamefont {Cottinet},\
		and\ \citenamefont {Guyomar}}]{Sebald_APL_101_022907}%
	\BibitemOpen
	\bibfield  {author} {\bibinfo {author} {\bibfnamefont {G.}~\bibnamefont
			{Sebald}}, \bibinfo {author} {\bibfnamefont {L.}~\bibnamefont {Seveyrat}},
		\bibinfo {author} {\bibfnamefont {J.-F.}\ \bibnamefont {Capsal}}, \bibinfo
		{author} {\bibfnamefont {P.-J.}\ \bibnamefont {Cottinet}},\ and\ \bibinfo
		{author} {\bibfnamefont {D.}~\bibnamefont {Guyomar}},\ }\bibfield  {title}
	{\enquote {\bibinfo {title} {Differential scanning calorimeter and infrared
				imaging for electrocaloric characterization of poly(vinylidene
				fluoride-trifluoroethylene-chlorofluoroethylene) terpolymer},}\ }\href@noop
	{} {\bibfield  {journal} {\bibinfo  {journal} {Appl. Phys. Lett.}\ }\textbf
		{\bibinfo {volume} {101}},\ \bibinfo {pages} {022907} (\bibinfo {year}
		{2012})}\BibitemShut {NoStop}%
	\bibitem [{\citenamefont {Guo}\ \emph {et~al.}(2014)\citenamefont {Guo},
		\citenamefont {Gao}, \citenamefont {Yu}, \citenamefont {Santhanam},
		\citenamefont {Fedder}, \citenamefont {McGaughey},\ and\ \citenamefont
		{Yao}}]{Guo_APL_105_031906}%
	\BibitemOpen
	\bibfield  {author} {\bibinfo {author} {\bibfnamefont {D.}~\bibnamefont
			{Guo}}, \bibinfo {author} {\bibfnamefont {J.}~\bibnamefont {Gao}}, \bibinfo
		{author} {\bibfnamefont {Y.-J.}\ \bibnamefont {Yu}}, \bibinfo {author}
		{\bibfnamefont {S.}~\bibnamefont {Santhanam}}, \bibinfo {author}
		{\bibfnamefont {G.~K.}\ \bibnamefont {Fedder}}, \bibinfo {author}
		{\bibfnamefont {A.~J.~H.}\ \bibnamefont {McGaughey}},\ and\ \bibinfo {author}
		{\bibfnamefont {S.~C.}\ \bibnamefont {Yao}},\ }\bibfield  {title} {\enquote
		{\bibinfo {title} {Electrocaloric characterization of a poly(vinylidene
				fluoride-trifluoroethylene-chlorofluoroethylene) terpolymer by infrared
				imaging},}\ }\href@noop {} {\bibfield  {journal} {\bibinfo  {journal} {Appl.
				Phys. Lett.}\ }\textbf {\bibinfo {volume} {105}},\ \bibinfo {pages} {031906}
		(\bibinfo {year} {2014})}\BibitemShut {NoStop}%
	\bibitem [{\citenamefont {Nouchokgwe}\ \emph {et~al.}(2021)\citenamefont
		{Nouchokgwe}, \citenamefont {Lheritier}, \citenamefont {Hong}, \citenamefont
		{Torell\'{o}}, \citenamefont {Faye}, \citenamefont {Jo}, \citenamefont
		{Bahl},\ and\ \citenamefont
		{Defay}}]{Nouchokgwe_NatureCommunications_12_3298}%
	\BibitemOpen
	\bibfield  {author} {\bibinfo {author} {\bibfnamefont {Y.}~\bibnamefont
			{Nouchokgwe}}, \bibinfo {author} {\bibfnamefont {P.}~\bibnamefont
			{Lheritier}}, \bibinfo {author} {\bibfnamefont {C.-H.}\ \bibnamefont {Hong}},
		\bibinfo {author} {\bibfnamefont {A.}~\bibnamefont {Torell\'{o}}}, \bibinfo
		{author} {\bibfnamefont {R.}~\bibnamefont {Faye}}, \bibinfo {author}
		{\bibfnamefont {W.}~\bibnamefont {Jo}}, \bibinfo {author} {\bibfnamefont
			{C.~R.~H.}\ \bibnamefont {Bahl}},\ and\ \bibinfo {author} {\bibfnamefont
			{E.}~\bibnamefont {Defay}},\ }\bibfield  {title} {\enquote {\bibinfo {title}
			{{Giant electrocaloric materials energy efficiency in highly ordered lead
					scandium tantalate}},}\ }\href@noop {} {\bibfield  {journal} {\bibinfo
			{journal} {Nature Commun.}\ }\textbf {\bibinfo {volume} {12}},\ \bibinfo
		{pages} {3298} (\bibinfo {year} {2021})}\BibitemShut {NoStop}%
	\bibitem [{\citenamefont {Nair}\ \emph {et~al.}(2019)\citenamefont {Nair},
		\citenamefont {Usui}, \citenamefont {Crossley}, \citenamefont {Kurdi},
		\citenamefont {Guzm\'{a}n-Verri}, \citenamefont {Moya}, \citenamefont
		{Hirose},\ and\ \citenamefont {Mathur}}]{Nair_Nature_575_468}%
	\BibitemOpen
	\bibfield  {author} {\bibinfo {author} {\bibfnamefont {B.}~\bibnamefont
			{Nair}}, \bibinfo {author} {\bibfnamefont {T.}~\bibnamefont {Usui}}, \bibinfo
		{author} {\bibfnamefont {S.}~\bibnamefont {Crossley}}, \bibinfo {author}
		{\bibfnamefont {S.}~\bibnamefont {Kurdi}}, \bibinfo {author} {\bibfnamefont
			{G.~G.}\ \bibnamefont {Guzm\'{a}n-Verri}}, \bibinfo {author} {\bibfnamefont
			{X.}~\bibnamefont {Moya}}, \bibinfo {author} {\bibfnamefont {S.}~\bibnamefont
			{Hirose}},\ and\ \bibinfo {author} {\bibfnamefont {N.~D.}\ \bibnamefont
			{Mathur}},\ }\bibfield  {title} {\enquote {\bibinfo {title} {{Large
					electrocaloric effects in oxide multilayer capacitors over a wide temperature
					range}},}\ }\href@noop {} {\bibfield  {journal} {\bibinfo  {journal}
			{Nature}\ }\textbf {\bibinfo {volume} {575}},\ \bibinfo {pages} {468--472}
		(\bibinfo {year} {2019})}\BibitemShut {NoStop}%
	\bibitem [{\citenamefont {Rowley}\ \emph {et~al.}(2015)\citenamefont {Rowley},
		\citenamefont {Hadjimichael}, \citenamefont {Ali}, \citenamefont {Durmaz},
		\citenamefont {Lashley}, \citenamefont {Cava},\ and\ \citenamefont
		{Scott}}]{Rowley_JPhysCondensMatter_27_395901}%
	\BibitemOpen
	\bibfield  {author} {\bibinfo {author} {\bibfnamefont {S.~E.}\ \bibnamefont
			{Rowley}}, \bibinfo {author} {\bibfnamefont {M.}~\bibnamefont
			{Hadjimichael}}, \bibinfo {author} {\bibfnamefont {M.~N.}\ \bibnamefont
			{Ali}}, \bibinfo {author} {\bibfnamefont {Y.~C.}\ \bibnamefont {Durmaz}},
		\bibinfo {author} {\bibfnamefont {J.~C.}\ \bibnamefont {Lashley}}, \bibinfo
		{author} {\bibfnamefont {R.~J.}\ \bibnamefont {Cava}},\ and\ \bibinfo
		{author} {\bibfnamefont {J.~F.}\ \bibnamefont {Scott}},\ }\bibfield  {title}
	{\enquote {\bibinfo {title} {Quantum criticality in a uniaxial organic
				ferroelectric},}\ }\href@noop {} {\bibfield  {journal} {\bibinfo  {journal}
			{J. Phys. Condens. Matter}\ }\textbf {\bibinfo {volume} {27}},\ \bibinfo
		{pages} {395901} (\bibinfo {year} {2015})}\BibitemShut {NoStop}%
	\bibitem [{\citenamefont {Brade\v{s}ko}\ \emph
		{et~al.}(2019{\natexlab{a}})\citenamefont {Brade\v{s}ko}, \citenamefont
		{Hedl}, \citenamefont {Fulanovi\'{c}}, \citenamefont {Novak},\ and\
		\citenamefont {Rojac}}]{Bradesko_APLMater_7_071111}%
	\BibitemOpen
	\bibfield  {author} {\bibinfo {author} {\bibfnamefont {A.}~\bibnamefont
			{Brade\v{s}ko}}, \bibinfo {author} {\bibfnamefont {A.}~\bibnamefont {Hedl}},
		\bibinfo {author} {\bibfnamefont {L.}~\bibnamefont {Fulanovi\'{c}}}, \bibinfo
		{author} {\bibfnamefont {N.}~\bibnamefont {Novak}},\ and\ \bibinfo {author}
		{\bibfnamefont {T.}~\bibnamefont {Rojac}},\ }\bibfield  {title} {\enquote
		{\bibinfo {title} {Self-heating of relaxor and ferroelectric ceramics during
				electrocaloric field cycling},}\ }\href@noop {} {\bibfield  {journal}
		{\bibinfo  {journal} {APL Mater.}\ }\textbf {\bibinfo {volume} {7}},\
		\bibinfo {pages} {071111} (\bibinfo {year} {2019}{\natexlab{a}})}\BibitemShut
	{NoStop}%
	\bibitem [{\citenamefont {Quintero}\ \emph {et~al.}(2011)\citenamefont
		{Quintero}, \citenamefont {Ghivelder}, \citenamefont {Gomez-Marlasca},\ and\
		\citenamefont {Parisi}}]{Quintero_APL_99_232908}%
	\BibitemOpen
	\bibfield  {author} {\bibinfo {author} {\bibfnamefont {M.}~\bibnamefont
			{Quintero}}, \bibinfo {author} {\bibfnamefont {L.}~\bibnamefont {Ghivelder}},
		\bibinfo {author} {\bibfnamefont {F.}~\bibnamefont {Gomez-Marlasca}},\ and\
		\bibinfo {author} {\bibfnamefont {F.}~\bibnamefont {Parisi}},\ }\bibfield
	{title} {\enquote {\bibinfo {title} {{Decoupling electrocaloric effect from
					Joule heating in a solid state cooling device}},}\ }\href@noop {} {\bibfield
		{journal} {\bibinfo  {journal} {Appl. Phys. Lett.}\ }\textbf {\bibinfo
			{volume} {99}},\ \bibinfo {pages} {232908} (\bibinfo {year}
		{2011})}\BibitemShut {NoStop}%
	\bibitem [{\citenamefont {D\"{o}ntgen}\ \emph {et~al.}(2015)\citenamefont
		{D\"{o}ntgen}, \citenamefont {Rudolph}, \citenamefont {Gottschall},
		\citenamefont {Gutfleisch}, \citenamefont {Salomon}, \citenamefont {Ludwig},\
		and\ \citenamefont {H\"{a}gele}}]{Doentgen_APL_106_032408}%
	\BibitemOpen
	\bibfield  {author} {\bibinfo {author} {\bibfnamefont {J.}~\bibnamefont
			{D\"{o}ntgen}}, \bibinfo {author} {\bibfnamefont {J.}~\bibnamefont
			{Rudolph}}, \bibinfo {author} {\bibfnamefont {T.}~\bibnamefont {Gottschall}},
		\bibinfo {author} {\bibfnamefont {O.}~\bibnamefont {Gutfleisch}}, \bibinfo
		{author} {\bibfnamefont {S.}~\bibnamefont {Salomon}}, \bibinfo {author}
		{\bibfnamefont {A.}~\bibnamefont {Ludwig}},\ and\ \bibinfo {author}
		{\bibfnamefont {D.}~\bibnamefont {H\"{a}gele}},\ }\bibfield  {title}
	{\enquote {\bibinfo {title} {{Temperature dependent low-field measurements of
					the magnetocaloric $\Delta T$ with sub-mK resolution in small volume and thin
					film samples}},}\ }\href@noop {} {\bibfield  {journal} {\bibinfo  {journal}
			{Appl. Phys. Lett.}\ }\textbf {\bibinfo {volume} {106}},\ \bibinfo {pages}
		{032408} (\bibinfo {year} {2015})}\BibitemShut {NoStop}%
	\bibitem [{\citenamefont {D\"{o}ntgen}\ \emph
		{et~al.}(2018{\natexlab{a}})\citenamefont {D\"{o}ntgen}, \citenamefont
		{Rudolph}, \citenamefont {Waske},\ and\ \citenamefont
		{H\"{a}gele}}]{Doentgen_RevSciInstrum_89_033909}%
	\BibitemOpen
	\bibfield  {author} {\bibinfo {author} {\bibfnamefont {J.}~\bibnamefont
			{D\"{o}ntgen}}, \bibinfo {author} {\bibfnamefont {J.}~\bibnamefont
			{Rudolph}}, \bibinfo {author} {\bibfnamefont {A.}~\bibnamefont {Waske}},\
		and\ \bibinfo {author} {\bibfnamefont {D.}~\bibnamefont {H\"{a}gele}},\
	}\bibfield  {title} {\enquote {\bibinfo {title} {{Modulation infrared
					thermometry of caloric effects at up to kHz frequencies}},}\ }\href@noop {}
	{\bibfield  {journal} {\bibinfo  {journal} {Rev. Sci. Instrum.}\ }\textbf
		{\bibinfo {volume} {89}},\ \bibinfo {pages} {033909} (\bibinfo {year}
		{2018}{\natexlab{a}})}\BibitemShut {NoStop}%
	\bibitem [{\citenamefont {D\"{o}ntgen}\ \emph
		{et~al.}(2018{\natexlab{b}})\citenamefont {D\"{o}ntgen}, \citenamefont
		{Rudolph}, \citenamefont {Gottschall}, \citenamefont {Gutfleisch},\ and\
		\citenamefont {H\"{a}gele}}]{Doentgen_EnergyTechnol_6_1470}%
	\BibitemOpen
	\bibfield  {author} {\bibinfo {author} {\bibfnamefont {J.}~\bibnamefont
			{D\"{o}ntgen}}, \bibinfo {author} {\bibfnamefont {J.}~\bibnamefont
			{Rudolph}}, \bibinfo {author} {\bibfnamefont {T.}~\bibnamefont {Gottschall}},
		\bibinfo {author} {\bibfnamefont {O.}~\bibnamefont {Gutfleisch}},\ and\
		\bibinfo {author} {\bibfnamefont {D.}~\bibnamefont {H\"{a}gele}},\ }\bibfield
	{title} {\enquote {\bibinfo {title} {{Millisecond Dynamics of the
					Magnetocaloric Effect in a First- and Second-Order Phase Transition
					Material}},}\ }\href@noop {} {\bibfield  {journal} {\bibinfo  {journal}
			{Energy Technol.}\ }\textbf {\bibinfo {volume} {6}},\ \bibinfo {pages} {1470}
		(\bibinfo {year} {2018}{\natexlab{b}})}\BibitemShut {NoStop}%
	\bibitem [{\citenamefont {Sawyer}\ and\ \citenamefont
		{Tower}(1930)}]{Sawyer_PR_35_269}%
	\BibitemOpen
	\bibfield  {author} {\bibinfo {author} {\bibfnamefont {C.~B.}\ \bibnamefont
			{Sawyer}}\ and\ \bibinfo {author} {\bibfnamefont {C.~H.}\ \bibnamefont
			{Tower}},\ }\bibfield  {title} {\enquote {\bibinfo {title} {{Rochelle Salt as
					a Dielectric}},}\ }\href@noop {} {\bibfield  {journal} {\bibinfo  {journal}
			{Phys. Rev.}\ }\textbf {\bibinfo {volume} {35}},\ \bibinfo {pages} {269--273}
		(\bibinfo {year} {1930})}\BibitemShut {NoStop}%
	\bibitem [{\citenamefont {Qian}\ \emph {et~al.}(2014)\citenamefont {Qian},
		\citenamefont {Ye}, \citenamefont {Zhang}, \citenamefont {Gu}, \citenamefont
		{Li}, \citenamefont {Randall},\ and\ \citenamefont
		{Zhang}}]{Qian_AdvFuncMater_24_1300}%
	\BibitemOpen
	\bibfield  {author} {\bibinfo {author} {\bibfnamefont {X.-S.}\ \bibnamefont
			{Qian}}, \bibinfo {author} {\bibfnamefont {H.-J.}\ \bibnamefont {Ye}},
		\bibinfo {author} {\bibfnamefont {Y.-T.}\ \bibnamefont {Zhang}}, \bibinfo
		{author} {\bibfnamefont {H.}~\bibnamefont {Gu}}, \bibinfo {author}
		{\bibfnamefont {X.}~\bibnamefont {Li}}, \bibinfo {author} {\bibfnamefont
			{C.~A.}\ \bibnamefont {Randall}},\ and\ \bibinfo {author} {\bibfnamefont
			{Q.~M.}\ \bibnamefont {Zhang}},\ }\bibfield  {title} {\enquote {\bibinfo
			{title} {{Giant Electrocaloric Response Over A Broad Temperature Range in
					Modified BaTiO$_3$ Ceramics}},}\ }\href@noop {} {\bibfield  {journal}
		{\bibinfo  {journal} {Adv. Funct. Mater.}\ }\textbf {\bibinfo {volume}
			{24}},\ \bibinfo {pages} {1300--1305} (\bibinfo {year} {2014})}\BibitemShut
	{NoStop}%
	\bibitem [{\citenamefont {Ma}\ \emph {et~al.}(2017{\natexlab{a}})\citenamefont
		{Ma}, \citenamefont {Molin}, \citenamefont {Shvartsman}, \citenamefont
		{Gebhardt}, \citenamefont {Lupascu}, \citenamefont {Albe},\ and\
		\citenamefont {Xu}}]{Ma_JAP_121_024103}%
	\BibitemOpen
	\bibfield  {author} {\bibinfo {author} {\bibfnamefont {Y.-B.}\ \bibnamefont
			{Ma}}, \bibinfo {author} {\bibfnamefont {C.}~\bibnamefont {Molin}}, \bibinfo
		{author} {\bibfnamefont {V.~V.}\ \bibnamefont {Shvartsman}}, \bibinfo
		{author} {\bibfnamefont {S.}~\bibnamefont {Gebhardt}}, \bibinfo {author}
		{\bibfnamefont {D.~C.}\ \bibnamefont {Lupascu}}, \bibinfo {author}
		{\bibfnamefont {K.}~\bibnamefont {Albe}},\ and\ \bibinfo {author}
		{\bibfnamefont {B.-X.}\ \bibnamefont {Xu}},\ }\bibfield  {title} {\enquote
		{\bibinfo {title} {{State transition and electrocaloric effect of
					BaZr$_x$Ti$_{1-x}$O$_3$: Simulation and experiment}},}\ }\href@noop {}
	{\bibfield  {journal} {\bibinfo  {journal} {J. Appl. Phys.}\ }\textbf
		{\bibinfo {volume} {121}},\ \bibinfo {pages} {024103} (\bibinfo {year}
		{2017}{\natexlab{a}})}\BibitemShut {NoStop}%
	\bibitem [{\citenamefont {Ye}\ \emph {et~al.}(2014)\citenamefont {Ye},
		\citenamefont {Qian}, \citenamefont {Jeong}, \citenamefont {Zhang},
		\citenamefont {Zhou}, \citenamefont {Shao}, \citenamefont {Zhen},\ and\
		\citenamefont {Zhang}}]{Ye_APL_105_152908}%
	\BibitemOpen
	\bibfield  {author} {\bibinfo {author} {\bibfnamefont {H.-J.}\ \bibnamefont
			{Ye}}, \bibinfo {author} {\bibfnamefont {X.-S.}\ \bibnamefont {Qian}},
		\bibinfo {author} {\bibfnamefont {D.-Y.}\ \bibnamefont {Jeong}}, \bibinfo
		{author} {\bibfnamefont {S.}~\bibnamefont {Zhang}}, \bibinfo {author}
		{\bibfnamefont {Y.}~\bibnamefont {Zhou}}, \bibinfo {author} {\bibfnamefont
			{W.-Z.}\ \bibnamefont {Shao}}, \bibinfo {author} {\bibfnamefont
			{L.}~\bibnamefont {Zhen}},\ and\ \bibinfo {author} {\bibfnamefont {Q.~M.}\
			\bibnamefont {Zhang}},\ }\bibfield  {title} {\enquote {\bibinfo {title}
			{{Giant electrocaloric effect in BaZr$_{0.2}$Ti$_{0.8}$O$_3$ thick film}},}\
	}\href@noop {} {\bibfield  {journal} {\bibinfo  {journal} {Appl. Phys.
				Lett.}\ }\textbf {\bibinfo {volume} {105}},\ \bibinfo {pages} {152908}
		(\bibinfo {year} {2014})}\BibitemShut {NoStop}%
	\bibitem [{\citenamefont {Sanlialp}\ \emph {et~al.}(2016)\citenamefont
		{Sanlialp}, \citenamefont {Molin}, \citenamefont {Shvartsman}, \citenamefont
		{Gebhardt},\ and\ \citenamefont
		{Lupascu}}]{Sanlialp_IEEETransUltrasonFerroelecFreqControl_63_1690}%
	\BibitemOpen
	\bibfield  {author} {\bibinfo {author} {\bibfnamefont {M.}~\bibnamefont
			{Sanlialp}}, \bibinfo {author} {\bibfnamefont {C.}~\bibnamefont {Molin}},
		\bibinfo {author} {\bibfnamefont {V.~V.}\ \bibnamefont {Shvartsman}},
		\bibinfo {author} {\bibfnamefont {S.}~\bibnamefont {Gebhardt}},\ and\
		\bibinfo {author} {\bibfnamefont {D.~C.}\ \bibnamefont {Lupascu}},\
	}\bibfield  {title} {\enquote {\bibinfo {title} {Modified differential
				scanning calorimeter for direct electrocaloric measurements},}\ }\href@noop
	{} {\bibfield  {journal} {\bibinfo  {journal} {IEEE Transactions on
				Ultrasonics, Ferroelectrics, and Frequency Control}\ }\textbf {\bibinfo
			{volume} {63}},\ \bibinfo {pages} {1690--1696} (\bibinfo {year}
		{2016})}\BibitemShut {NoStop}%
	\bibitem [{\citenamefont {Sanlialp}\ \emph {et~al.}(2018)\citenamefont
		{Sanlialp}, \citenamefont {Shvartsman}, \citenamefont {Faye}, \citenamefont
		{Karabasov}, \citenamefont {Molin}, \citenamefont {Gebhardt}, \citenamefont
		{Defay},\ and\ \citenamefont {Lupascu}}]{Sanlialp_RSI_89_034903}%
	\BibitemOpen
	\bibfield  {author} {\bibinfo {author} {\bibfnamefont {M.}~\bibnamefont
			{Sanlialp}}, \bibinfo {author} {\bibfnamefont {V.~V.}\ \bibnamefont
			{Shvartsman}}, \bibinfo {author} {\bibfnamefont {R.}~\bibnamefont {Faye}},
		\bibinfo {author} {\bibfnamefont {M.~O.}\ \bibnamefont {Karabasov}}, \bibinfo
		{author} {\bibfnamefont {C.}~\bibnamefont {Molin}}, \bibinfo {author}
		{\bibfnamefont {S.}~\bibnamefont {Gebhardt}}, \bibinfo {author}
		{\bibfnamefont {E.}~\bibnamefont {Defay}},\ and\ \bibinfo {author}
		{\bibfnamefont {D.~C.}\ \bibnamefont {Lupascu}},\ }\bibfield  {title}
	{\enquote {\bibinfo {title} {Quasi-adiabatic calorimeter for direct
				electrocaloric measurements},}\ }\href@noop {} {\bibfield  {journal}
		{\bibinfo  {journal} {Rev. Sci. Inst.}\ }\textbf {\bibinfo {volume} {89}},\
		\bibinfo {pages} {034903} (\bibinfo {year} {2018})}\BibitemShut {NoStop}%
	\bibitem [{Note1()}]{Note1}%
	\BibitemOpen
	\bibinfo {note} {The data are corrected for the measured frequency response
		of the IR detector amplifier.}\BibitemShut {Stop}%
	\bibitem [{\citenamefont {Molin}\ and\ \citenamefont
		{Gebhardt}(2016)}]{Molin_Ferroelectrics_498_111}%
	\BibitemOpen
	\bibfield  {author} {\bibinfo {author} {\bibfnamefont {C.}~\bibnamefont
			{Molin}}\ and\ \bibinfo {author} {\bibfnamefont {S.}~\bibnamefont
			{Gebhardt}},\ }\bibfield  {title} {\enquote {\bibinfo {title} {{PMN--8PT
					device structures for electrocaloric cooling applications}},}\ }\href@noop {}
	{\bibfield  {journal} {\bibinfo  {journal} {Ferroelectrics}\ }\textbf
		{\bibinfo {volume} {498}},\ \bibinfo {pages} {111--119} (\bibinfo {year}
		{2016})}\BibitemShut {NoStop}%
	\bibitem [{Note2()}]{Note2}%
	\BibitemOpen
	\bibinfo {note} {The layer thickness of the paint was determined by a laser
		scanning microscope.}\BibitemShut {Stop}%
	\bibitem [{\citenamefont {Molin}\ \emph {et~al.}(2017)\citenamefont {Molin},
		\citenamefont {Per\"{a}ntie}, \citenamefont {Goupil}, \citenamefont
		{Weyland}, \citenamefont {Sanlialp}, \citenamefont {Stingelin}, \citenamefont
		{Novak}, \citenamefont {Lupascu},\ and\ \citenamefont
		{Gebhardt}}]{Molin_JAmCeramSoc_100_2885}%
	\BibitemOpen
	\bibfield  {author} {\bibinfo {author} {\bibfnamefont {C.}~\bibnamefont
			{Molin}}, \bibinfo {author} {\bibfnamefont {J.}~\bibnamefont {Per\"{a}ntie}},
		\bibinfo {author} {\bibfnamefont {F.~L.}\ \bibnamefont {Goupil}}, \bibinfo
		{author} {\bibfnamefont {F.}~\bibnamefont {Weyland}}, \bibinfo {author}
		{\bibfnamefont {M.}~\bibnamefont {Sanlialp}}, \bibinfo {author}
		{\bibfnamefont {N.}~\bibnamefont {Stingelin}}, \bibinfo {author}
		{\bibfnamefont {N.}~\bibnamefont {Novak}}, \bibinfo {author} {\bibfnamefont
			{D.~C.}\ \bibnamefont {Lupascu}},\ and\ \bibinfo {author} {\bibfnamefont
			{S.}~\bibnamefont {Gebhardt}},\ }\bibfield  {title} {\enquote {\bibinfo
			{title} {{Comparison of direct electrocaloric characterization methods
					exemplified by 0.92 Pb(Mg$_{1/3}$Nb$_{2/3}$)O$_3$-0.08 PbTiO$_3$ multilayer
					ceramics}},}\ }\href@noop {} {\bibfield  {journal} {\bibinfo  {journal} {J.
				Am. Ceram. Soc.}\ }\textbf {\bibinfo {volume} {100}},\ \bibinfo {pages}
		{2885--2892} (\bibinfo {year} {2017})}\BibitemShut {NoStop}%
	\bibitem [{\citenamefont {Molin}\ \emph {et~al.}(2015)\citenamefont {Molin},
		\citenamefont {Sanlialp}, \citenamefont {Shvartsman}, \citenamefont
		{Lupascu}, \citenamefont {Neumeister}, \citenamefont {Sch\"{o}necker},\ and\
		\citenamefont {Gebhardt}}]{Molin_JEurCeramSoc_35_2065}%
	\BibitemOpen
	\bibfield  {author} {\bibinfo {author} {\bibfnamefont {C.}~\bibnamefont
			{Molin}}, \bibinfo {author} {\bibfnamefont {M.}~\bibnamefont {Sanlialp}},
		\bibinfo {author} {\bibfnamefont {V.}~\bibnamefont {Shvartsman}}, \bibinfo
		{author} {\bibfnamefont {D.}~\bibnamefont {Lupascu}}, \bibinfo {author}
		{\bibfnamefont {P.}~\bibnamefont {Neumeister}}, \bibinfo {author}
		{\bibfnamefont {A.}~\bibnamefont {Sch\"{o}necker}},\ and\ \bibinfo {author}
		{\bibfnamefont {S.}~\bibnamefont {Gebhardt}},\ }\bibfield  {title} {\enquote
		{\bibinfo {title} {{Effect of dopants on the electrocaloric effect of 0.92
					Pb(Mg$_{1/3}$Nb$_{2/3}$)O$_3$-0.08 PbTiO$_3$ ceramics}},}\ }\href@noop {}
	{\bibfield  {journal} {\bibinfo  {journal} {J. Eur. Ceram. Soc.}\ }\textbf
		{\bibinfo {volume} {35}},\ \bibinfo {pages} {2065--2071} (\bibinfo {year}
		{2015})}\BibitemShut {NoStop}%
	\bibitem [{\citenamefont {Molin}\ \emph {et~al.}(2018)\citenamefont {Molin},
		\citenamefont {Neumeister}, \citenamefont {Neubert},\ and\ \citenamefont
		{Gebhardt}}]{Molin_EnergyTechnol_6_1543}%
	\BibitemOpen
	\bibfield  {author} {\bibinfo {author} {\bibfnamefont {C.}~\bibnamefont
			{Molin}}, \bibinfo {author} {\bibfnamefont {P.}~\bibnamefont {Neumeister}},
		\bibinfo {author} {\bibfnamefont {H.}~\bibnamefont {Neubert}},\ and\ \bibinfo
		{author} {\bibfnamefont {S.~E.}\ \bibnamefont {Gebhardt}},\ }\bibfield
	{title} {\enquote {\bibinfo {title} {{Multilayer Ceramics for Electrocaloric
					Cooling Applications}},}\ }\href@noop {} {\bibfield  {journal} {\bibinfo
			{journal} {Energy Technol.}\ }\textbf {\bibinfo {volume} {6}},\ \bibinfo
		{pages} {1543--1552} (\bibinfo {year} {2018})}\BibitemShut {NoStop}%
	\bibitem [{\citenamefont {Greiner}\ \emph {et~al.}(2018)\citenamefont
		{Greiner}, \citenamefont {Molin}, \citenamefont {Neubert}, \citenamefont
		{Gebhardt},\ and\ \citenamefont {Neumeister}}]{Greiner_EnergyTechnol_6_1535}%
	\BibitemOpen
	\bibfield  {author} {\bibinfo {author} {\bibfnamefont {A.}~\bibnamefont
			{Greiner}}, \bibinfo {author} {\bibfnamefont {C.}~\bibnamefont {Molin}},
		\bibinfo {author} {\bibfnamefont {H.}~\bibnamefont {Neubert}}, \bibinfo
		{author} {\bibfnamefont {S.~E.}\ \bibnamefont {Gebhardt}},\ and\ \bibinfo
		{author} {\bibfnamefont {P.}~\bibnamefont {Neumeister}},\ }\bibfield  {title}
	{\enquote {\bibinfo {title} {{Direct Measurement of the Electrocaloric
					Temperature Change in Multilayer Ceramic Components using Resistance-Welded
					Thermocouple Wires}},}\ }\href@noop {} {\bibfield  {journal} {\bibinfo
			{journal} {Energy Technol.}\ }\textbf {\bibinfo {volume} {6}},\ \bibinfo
		{pages} {1535--1542} (\bibinfo {year} {2018})}\BibitemShut {NoStop}%
	\bibitem [{\citenamefont {Ma}\ \emph {et~al.}(2017{\natexlab{b}})\citenamefont
		{Ma}, \citenamefont {Zhang}, \citenamefont {Tong}, \citenamefont {Huber},
		\citenamefont {Kornbluh}, \citenamefont {Ju},\ and\ \citenamefont
		{Pei}}]{Ma_Science_357_1130}%
	\BibitemOpen
	\bibfield  {author} {\bibinfo {author} {\bibfnamefont {R.}~\bibnamefont
			{Ma}}, \bibinfo {author} {\bibfnamefont {Z.}~\bibnamefont {Zhang}}, \bibinfo
		{author} {\bibfnamefont {K.}~\bibnamefont {Tong}}, \bibinfo {author}
		{\bibfnamefont {D.}~\bibnamefont {Huber}}, \bibinfo {author} {\bibfnamefont
			{R.}~\bibnamefont {Kornbluh}}, \bibinfo {author} {\bibfnamefont {Y.~S.}\
			\bibnamefont {Ju}},\ and\ \bibinfo {author} {\bibfnamefont {Q.}~\bibnamefont
			{Pei}},\ }\bibfield  {title} {\enquote {\bibinfo {title} {Highly efficient
				electrocaloric cooling with electrostatic actuation},}\ }\href@noop {}
	{\bibfield  {journal} {\bibinfo  {journal} {Science}\ }\textbf {\bibinfo
			{volume} {357}},\ \bibinfo {pages} {1130--1134} (\bibinfo {year}
		{2017}{\natexlab{b}})}\BibitemShut {NoStop}%
	\bibitem [{\citenamefont {Neese}\ \emph {et~al.}(2009)\citenamefont {Neese},
		\citenamefont {Lu}, \citenamefont {Chu},\ and\ \citenamefont
		{Zhang}}]{Neese_APL_94_042910}%
	\BibitemOpen
	\bibfield  {author} {\bibinfo {author} {\bibfnamefont {B.}~\bibnamefont
			{Neese}}, \bibinfo {author} {\bibfnamefont {S.~G.}\ \bibnamefont {Lu}},
		\bibinfo {author} {\bibfnamefont {B.}~\bibnamefont {Chu}},\ and\ \bibinfo
		{author} {\bibfnamefont {Q.~M.}\ \bibnamefont {Zhang}},\ }\bibfield  {title}
	{\enquote {\bibinfo {title} {Electrocaloric effect of the relaxor
				ferroelectric poly(vinylidene
				fluoride-trifluoroethylene-chlorofluoroethylene) terpolymer},}\ }\href@noop
	{} {\bibfield  {journal} {\bibinfo  {journal} {Appl. Phys. Lett.}\ }\textbf
		{\bibinfo {volume} {94}},\ \bibinfo {pages} {042910} (\bibinfo {year}
		{2009})}\BibitemShut {NoStop}%
	\bibitem [{\citenamefont {Hambal}\ \emph {et~al.}(2021)\citenamefont {Hambal},
		\citenamefont {Shvartsman}, \citenamefont {Lewin}, \citenamefont {Huat},
		\citenamefont {Chen}, \citenamefont {Michiels}, \citenamefont {Zhang},\ and\
		\citenamefont {Lupascu}}]{Hambal_Polymers_13_1343}%
	\BibitemOpen
	\bibfield  {author} {\bibinfo {author} {\bibfnamefont {Y.}~\bibnamefont
			{Hambal}}, \bibinfo {author} {\bibfnamefont {V.}~\bibnamefont {Shvartsman}},
		\bibinfo {author} {\bibfnamefont {D.}~\bibnamefont {Lewin}}, \bibinfo
		{author} {\bibfnamefont {C.}~\bibnamefont {Huat}}, \bibinfo {author}
		{\bibfnamefont {X.}~\bibnamefont {Chen}}, \bibinfo {author} {\bibfnamefont
			{I.}~\bibnamefont {Michiels}}, \bibinfo {author} {\bibfnamefont
			{Q.}~\bibnamefont {Zhang}},\ and\ \bibinfo {author} {\bibfnamefont
			{D.}~\bibnamefont {Lupascu}},\ }\bibfield  {title} {\enquote {\bibinfo
			{title} {{Effect of Composition on Polarization Hysteresis and Energy Storage
					Ability of P(VDF-TrFE-CFE) Relaxor Terpolymers}},}\ }\href@noop {} {\bibfield
		{journal} {\bibinfo  {journal} {Polymers}\ }\textbf {\bibinfo {volume}
			{13}},\ \bibinfo {pages} {1343} (\bibinfo {year} {2021})}\BibitemShut
	{NoStop}%
	\bibitem [{\citenamefont {Li}\ \emph {et~al.}(2011)\citenamefont {Li},
		\citenamefont {Qian}, \citenamefont {Lu}, \citenamefont {Cheng},
		\citenamefont {Fang},\ and\ \citenamefont {Zhang}}]{Li_APL_99_052907}%
	\BibitemOpen
	\bibfield  {author} {\bibinfo {author} {\bibfnamefont {X.}~\bibnamefont
			{Li}}, \bibinfo {author} {\bibfnamefont {X.-s.}\ \bibnamefont {Qian}},
		\bibinfo {author} {\bibfnamefont {S.~G.}\ \bibnamefont {Lu}}, \bibinfo
		{author} {\bibfnamefont {J.}~\bibnamefont {Cheng}}, \bibinfo {author}
		{\bibfnamefont {Z.}~\bibnamefont {Fang}},\ and\ \bibinfo {author}
		{\bibfnamefont {Q.~M.}\ \bibnamefont {Zhang}},\ }\bibfield  {title} {\enquote
		{\bibinfo {title} {Tunable temperature dependence of electrocaloric effect in
				ferroelectric relaxor poly(vinylidene
				fluoride-trifluoroethylene-chlorofluoroethylene terpolymer},}\ }\href@noop {}
	{\bibfield  {journal} {\bibinfo  {journal} {Appl. Phys. Lett.}\ }\textbf
		{\bibinfo {volume} {99}},\ \bibinfo {pages} {052907} (\bibinfo {year}
		{2011})}\BibitemShut {NoStop}%
	\bibitem [{\citenamefont {Liu}\ \emph {et~al.}(2019)\citenamefont {Liu},
		\citenamefont {Zhang}, \citenamefont {Haibibu}, \citenamefont {Han},\ and\
		\citenamefont {Wang}}]{Liu_JAP_126_234102}%
	\BibitemOpen
	\bibfield  {author} {\bibinfo {author} {\bibfnamefont {Y.}~\bibnamefont
			{Liu}}, \bibinfo {author} {\bibfnamefont {G.}~\bibnamefont {Zhang}}, \bibinfo
		{author} {\bibfnamefont {A.}~\bibnamefont {Haibibu}}, \bibinfo {author}
		{\bibfnamefont {Z.}~\bibnamefont {Han}},\ and\ \bibinfo {author}
		{\bibfnamefont {Q.}~\bibnamefont {Wang}},\ }\bibfield  {title} {\enquote
		{\bibinfo {title} {High cyclic stability of electrocaloric effect in relaxor
				poly(vinylidene fluoride-trifluoroethylene-chlorofluoroethylene) terpolymers
				in the absence of ferroelectric phase transition},}\ }\href@noop {}
	{\bibfield  {journal} {\bibinfo  {journal} {J. Appl. Phys.}\ }\textbf
		{\bibinfo {volume} {126}},\ \bibinfo {pages} {234102} (\bibinfo {year}
		{2019})}\BibitemShut {NoStop}%
	\bibitem [{\citenamefont {Jia}\ and\ \citenamefont
		{Sungtaek~Ju}(2013)}]{Jia_APL_103_042903}%
	\BibitemOpen
	\bibfield  {author} {\bibinfo {author} {\bibfnamefont {Y.}~\bibnamefont
			{Jia}}\ and\ \bibinfo {author} {\bibfnamefont {Y.}~\bibnamefont
			{Sungtaek~Ju}},\ }\bibfield  {title} {\enquote {\bibinfo {title} {Direct
				characterization of the electrocaloric effects in thin films supported on
				substrates},}\ }\href@noop {} {\bibfield  {journal} {\bibinfo  {journal}
			{Appl. Phys. Lett.}\ }\textbf {\bibinfo {volume} {103}},\ \bibinfo {pages}
		{042903} (\bibinfo {year} {2013})}\BibitemShut {NoStop}%
	\bibitem [{\citenamefont
		{Lupascu}(2004)}]{LupascuFatigueFerroelectricCeramics}%
	\BibitemOpen
	\bibfield  {author} {\bibinfo {author} {\bibfnamefont {D.~C.}\ \bibnamefont
			{Lupascu}},\ }\href@noop {} {\emph {\bibinfo {title} {{Fatigue in
					Ferroelectric Ceramics and Related Issues}}}}\ (\bibinfo  {publisher}
	{{Springer}},\ \bibinfo {year} {2004})\BibitemShut {NoStop}%
	\bibitem [{\citenamefont {Lupascu}\ and\ \citenamefont
		{R\"{o}del}(2005)}]{Lupascu_AdvEngMater_7_882}%
	\BibitemOpen
	\bibfield  {author} {\bibinfo {author} {\bibfnamefont {D.}~\bibnamefont
			{Lupascu}}\ and\ \bibinfo {author} {\bibfnamefont {J.}~\bibnamefont
			{R\"{o}del}},\ }\bibfield  {title} {\enquote {\bibinfo {title} {{Fatigue In
					Bulk Lead Zirconate Titanate Actuator Materials}},}\ }\href@noop {}
	{\bibfield  {journal} {\bibinfo  {journal} {Adv. Eng. Mater.}\ }\textbf
		{\bibinfo {volume} {7}},\ \bibinfo {pages} {882--898} (\bibinfo {year}
		{2005})}\BibitemShut {NoStop}%
	\bibitem [{\citenamefont {Fulanovi\'{c}}\ \emph {et~al.}(2017)\citenamefont
		{Fulanovi\'{c}}, \citenamefont {Koruza}, \citenamefont {Novak}, \citenamefont
		{Weyland}, \citenamefont {Mali\v{c}},\ and\ \citenamefont
		{Bobnar}}]{Fulanovic_JEurCeramSoc_37_5105}%
	\BibitemOpen
	\bibfield  {author} {\bibinfo {author} {\bibfnamefont {L.}~\bibnamefont
			{Fulanovi\'{c}}}, \bibinfo {author} {\bibfnamefont {J.}~\bibnamefont
			{Koruza}}, \bibinfo {author} {\bibfnamefont {N.}~\bibnamefont {Novak}},
		\bibinfo {author} {\bibfnamefont {F.}~\bibnamefont {Weyland}}, \bibinfo
		{author} {\bibfnamefont {B.}~\bibnamefont {Mali\v{c}}},\ and\ \bibinfo
		{author} {\bibfnamefont {V.}~\bibnamefont {Bobnar}},\ }\bibfield  {title}
	{\enquote {\bibinfo {title} {{Fatigue-less electrocaloric effect in relaxor
					Pb(Mg$_{1/3}$Nb$_{2/3}$)O$_3$ multilayer elements}},}\ }\href@noop {}
	{\bibfield  {journal} {\bibinfo  {journal} {J. Eur. Ceram. Soc.}\ }\textbf
		{\bibinfo {volume} {37}},\ \bibinfo {pages} {5105--5108} (\bibinfo {year}
		{2017})}\BibitemShut {NoStop}%
	\bibitem [{\citenamefont {Weyland}\ \emph {et~al.}(2018)\citenamefont
		{Weyland}, \citenamefont {Eisele}, \citenamefont {Steiner}, \citenamefont
		{Frömling}, \citenamefont {Rossetti}, \citenamefont {R\"{o}del},\ and\
		\citenamefont {Novak}}]{Weyland_JEuropCeramSoc_38_551}%
	\BibitemOpen
	\bibfield  {author} {\bibinfo {author} {\bibfnamefont {F.}~\bibnamefont
			{Weyland}}, \bibinfo {author} {\bibfnamefont {T.}~\bibnamefont {Eisele}},
		\bibinfo {author} {\bibfnamefont {S.}~\bibnamefont {Steiner}}, \bibinfo
		{author} {\bibfnamefont {T.}~\bibnamefont {Frömling}}, \bibinfo {author}
		{\bibfnamefont {G.~A.}\ \bibnamefont {Rossetti}}, \bibinfo {author}
		{\bibfnamefont {J.}~\bibnamefont {R\"{o}del}},\ and\ \bibinfo {author}
		{\bibfnamefont {N.}~\bibnamefont {Novak}},\ }\bibfield  {title} {\enquote
		{\bibinfo {title} {Long term stability of electrocaloric response in barium
				zirconate titanate},}\ }\href@noop {} {\bibfield  {journal} {\bibinfo
			{journal} {J. Eur. Ceram. Soc.}\ }\textbf {\bibinfo {volume} {38}},\ \bibinfo
		{pages} {551--556} (\bibinfo {year} {2018})}\BibitemShut {NoStop}%
	\bibitem [{\citenamefont {Brade\v{s}ko}\ \emph
		{et~al.}(2019{\natexlab{b}})\citenamefont {Brade\v{s}ko}, \citenamefont
		{Fulanovi\'{c}}, \citenamefont {Vrabelj}, \citenamefont {Otoni\v{c}ar},
		\citenamefont {Ur\v{s}i\v{c}}, \citenamefont {Henriques}, \citenamefont
		{Chung}, \citenamefont {Jones}, \citenamefont {Mali\v{c}}, \citenamefont
		{Kutnjak},\ and\ \citenamefont {Rojac}}]{Bradesko_ActaMaterialia_169_275}%
	\BibitemOpen
	\bibfield  {author} {\bibinfo {author} {\bibfnamefont {A.}~\bibnamefont
			{Brade\v{s}ko}}, \bibinfo {author} {\bibfnamefont {L.}~\bibnamefont
			{Fulanovi\'{c}}}, \bibinfo {author} {\bibfnamefont {M.}~\bibnamefont
			{Vrabelj}}, \bibinfo {author} {\bibfnamefont {M.}~\bibnamefont
			{Otoni\v{c}ar}}, \bibinfo {author} {\bibfnamefont {H.}~\bibnamefont
			{Ur\v{s}i\v{c}}}, \bibinfo {author} {\bibfnamefont {A.}~\bibnamefont
			{Henriques}}, \bibinfo {author} {\bibfnamefont {C.-C.}\ \bibnamefont
			{Chung}}, \bibinfo {author} {\bibfnamefont {J.~L.}\ \bibnamefont {Jones}},
		\bibinfo {author} {\bibfnamefont {B.}~\bibnamefont {Mali\v{c}}}, \bibinfo
		{author} {\bibfnamefont {Z.}~\bibnamefont {Kutnjak}},\ and\ \bibinfo {author}
		{\bibfnamefont {T.}~\bibnamefont {Rojac}},\ }\bibfield  {title} {\enquote
		{\bibinfo {title} {Electrocaloric fatigue of lead magnesium niobate mediated
				by an electric-field-induced phase transformation},}\ }\href@noop {}
	{\bibfield  {journal} {\bibinfo  {journal} {Acta Materialia}\ }\textbf
		{\bibinfo {volume} {169}},\ \bibinfo {pages} {275--283} (\bibinfo {year}
		{2019}{\natexlab{b}})}\BibitemShut {NoStop}%
	\bibitem [{\citenamefont {Del~Duca}\ \emph {et~al.}(2020)\citenamefont
		{Del~Duca}, \citenamefont {Tu\v{s}ek}, \citenamefont {Maiorino},
		\citenamefont {Fulanovi\'{c}}, \citenamefont {Brade\v{s}ko}, \citenamefont
		{Plaznik}, \citenamefont {Mali\v{c}}, \citenamefont {Aprea},\ and\
		\citenamefont {Kitanovski}}]{delDuca_JAP_128_104102}%
	\BibitemOpen
	\bibfield  {author} {\bibinfo {author} {\bibfnamefont {M.~G.}\ \bibnamefont
			{Del~Duca}}, \bibinfo {author} {\bibfnamefont {J.}~\bibnamefont {Tu\v{s}ek}},
		\bibinfo {author} {\bibfnamefont {A.}~\bibnamefont {Maiorino}}, \bibinfo
		{author} {\bibfnamefont {L.}~\bibnamefont {Fulanovi\'{c}}}, \bibinfo {author}
		{\bibfnamefont {A.}~\bibnamefont {Brade\v{s}ko}}, \bibinfo {author}
		{\bibfnamefont {U.}~\bibnamefont {Plaznik}}, \bibinfo {author} {\bibfnamefont
			{B.}~\bibnamefont {Mali\v{c}}}, \bibinfo {author} {\bibfnamefont
			{C.}~\bibnamefont {Aprea}},\ and\ \bibinfo {author} {\bibfnamefont
			{A.}~\bibnamefont {Kitanovski}},\ }\bibfield  {title} {\enquote {\bibinfo
			{title} {{Comprehensive evaluation of electrocaloric effect and fatigue
					behavior in the 0.9Pb(Mg$_{1/3}$Nb$_{2/3}$)O3-0.1PbTiO$_3$ bulk relaxor
					ferroelectric ceramic}},}\ }\href@noop {} {\bibfield  {journal} {\bibinfo
			{journal} {J. Appl. Phys.}\ }\textbf {\bibinfo {volume} {128}},\ \bibinfo
		{pages} {104102} (\bibinfo {year} {2020})}\BibitemShut {NoStop}%
	\bibitem [{\citenamefont {Nuffer}, \citenamefont {Lupascu},\ and\ \citenamefont
		{R\"{o}del}(2000)}]{Nuffer_ActaMaterialia_48_3783}%
	\BibitemOpen
	\bibfield  {author} {\bibinfo {author} {\bibfnamefont {J.}~\bibnamefont
			{Nuffer}}, \bibinfo {author} {\bibfnamefont {D.~C.}\ \bibnamefont
			{Lupascu}},\ and\ \bibinfo {author} {\bibfnamefont {J.}~\bibnamefont
			{R\"{o}del}},\ }\bibfield  {title} {\enquote {\bibinfo {title} {{Damage
					evolution in ferroelectric PZT induced by bipolar electric cycling}},}\
	}\href@noop {} {\bibfield  {journal} {\bibinfo  {journal} {Acta Materialia}\
		}\textbf {\bibinfo {volume} {48}},\ \bibinfo {pages} {3783--3794} (\bibinfo
		{year} {2000})}\BibitemShut {NoStop}%
	\bibitem [{\citenamefont {Lupascu}, \citenamefont {Granzow},\ and\
		\citenamefont {Woike}(2004)}]{Lupascu_EPL_68_733}%
	\BibitemOpen
	\bibfield  {author} {\bibinfo {author} {\bibfnamefont {D.~C.}\ \bibnamefont
			{Lupascu}}, \bibinfo {author} {\bibfnamefont {T.}~\bibnamefont {Granzow}},\
		and\ \bibinfo {author} {\bibfnamefont {T.}~\bibnamefont {Woike}},\ }\bibfield
	{title} {\enquote {\bibinfo {title} {{Discontinuous domain wall motion in
					the relaxor ferroelectric Sr$_{0.61}$Ba$_{0.39}$Nb$_2$O$_6$}},}\ }\href@noop
	{} {\bibfield  {journal} {\bibinfo  {journal} {Europhys. Lett.}\ }\textbf
		{\bibinfo {volume} {68}},\ \bibinfo {pages} {733--739} (\bibinfo {year}
		{2004})}\BibitemShut {NoStop}%
	\bibitem [{\citenamefont {Fatuzzo}\ and\ \citenamefont
		{Merz}(1967)}]{FatuzzoMerzFerroelectricity}%
	\BibitemOpen
	\bibfield  {author} {\bibinfo {author} {\bibfnamefont {F.}~\bibnamefont
			{Fatuzzo}}\ and\ \bibinfo {author} {\bibfnamefont {W.~J.}\ \bibnamefont
			{Merz}},\ }\href@noop {} {\emph {\bibinfo {title} {Ferroelectricity}}}\
	(\bibinfo  {publisher} {North-Holland Pubishing Company, Amsterdam},\
	\bibinfo {year} {1967})\BibitemShut {NoStop}%
\end{thebibliography}
%

\end{document}